\documentclass[usenatbib, twocolumn, nofootinbib]{aastex631}
\usepackage{amsmath}
\usepackage{commands}
\usepackage{wrapfig}
\usepackage{comment}
\usepackage{multirow}
\usepackage{soul}
\usepackage{wrapfig}
\usepackage[normalem]{ulem}
\usepackage[mathlines]{lineno}
\newcommand{\snr}{S/N}
\newcommand{\refresponse}[1]{{#1}}

\newcommand{\kp}[1]{\textcolor{blue}{[KP: #1]}}

\newcommand{\lloyd}[1]{{\color{magenta}#1}}
\newcommand{\commenter}[1]{{}}
\newcommand{\specialparaheader}[1]{\paragraph{{\it{#1}}}}

\begin{document}
%paremeter constraints tables

%%%%%%%%%%%%%%%%%%%%%%%%%%%%%%%%%%%%%%%%%%%%%%%%%%%%%%%%%%%%%%%%%%%%%%%%%%%%%%%%
%%%%%%%%%%%%%%%%%%%%%%%%%%%%%%%%%%%%%%%%%%%%%%%%%%%%%%%%%%%%%%%%%%%%%%%%%%%%%%%%
%%%%%%%%%%%%%%%%%%%%%%%%%%%%%%%%%%%%%%%%%%%%%%%%%%%%%%%%%%%%%%%%%%%%%%%%%%%%%%%%
%LCDM
%\raggedbottom
\newcommand{\lcdmtable}{
\begin{deluxetable*}{lccccc}
%\tabletypesize{\small}
%\tablewidth{0.5\textwidth}
%\def\arraystretch{1.5}
%\renewcommand{\arraystretch}{1.}
\tablecaption{Forecasted $1\sigma$ errors on \llcdm\ parameters for various survey regions using SPT data. The numbers in the parentheses correspond to the constraints obtained when \sptnew{} is combined with \planck. Note that when \sptnew{} is combined with \planck, we remove information below $\ell_{\rm min} = 2000, 1000, 750, 30$ for $TT, TE, EE$ and $\phi\phi$ power spectra respectively from \sptnew, to avoid double counting information.}
\label{tab_lcdm_param_table}
\tablehead{
 Parameter &           \mainfield{} &           \summerfield{} &   \extfield{} &             \mainplussummerfield{} &               \allspt{}
}
\startdata
\hline \hline
$A_s$ \hfill [$10^{-11}$]\hfill & 2.79 (2.47) & 2.66 (2.50) & 2.54 (2.42) & 2.53 (2.38) & 2.35 (2.29) \\
$H_0 [{\rm km/s/Mpc}]$ \hfill [$10^{-1}$]\hfill  & 3.28 (3.16) & 3.91 (3.47) & 3.49 (3.26) & 3.03 (2.95) & 2.78 (2.78)\\ 
$n_s$ \hfill [$10^{-3}$]\hfill  & 7.23 (3.14) & 6.12 (3.16) & 5.10 (3.03) & 4.76 (3.00) & 3.66 (2.83) \\
$\ombh$ \hfill[$10^{-5}$]\hfill & 12.9 (7.87) & 14.9 (8.68) & 11.9 (7.66) & 9.5 (6.7) & 7.37 (5.67)\\     
$\omch$\hfill [$10^{-4}$]\hfill & 8.86 (7.71) & 9.84 (8.34) & 8.92 (7.97) & 8.07 (7.39) & 7.36 (7.10)\\
$\taure$ \hfill [$10^{-3}$]\hfill & 6.85 (6.64) & 6.78 (6.65) & 6.59 (6.47) & 6.63 (6.42) & 6.28 (6.19)
\enddata
\end{deluxetable*}
}

\newcommand{\lcdmtablebothmuseandfisher}{
\begin{deluxetable*}{l|c|c|c|c|c|c|c|c|c|c}
\tabletypesize{\small}
%\tablewidth{0.5\textwidth}
%\def\arraystretch{1.2}
%\renewcommand{\arraystretch}{1.}
\tablecaption{Forecasted 1$\sigma$ errors on \llcdm\ parameters for various \sptnew{} survey regions. We report the constraints from both MUSE as well as the standard Fisher formalism. The two methods agree reasonably well.}
\label{tab_ext_param_table}
\tablehead{
\multirow{2}{*}{Parameter} & \multicolumn{2}{c|}{\mainfield} & \multicolumn{2}{c|}{\summerfield} &   \multicolumn{2}{c|}{\extfield} & \multicolumn{2}{c|}{\mainplussummerfield} & \multicolumn{2}{c}{\allspt} \\
 \cline{2-11}
& MUSE & Fisher & MUSE & Fisher & MUSE & Fisher & MUSE & Fisher & MUSE & Fisher
}
\startdata
\hline \hline
$A_{s}$ \hfill [$10^{-11}$] & 2.78 & --- & 2.73 & --- & 2.53 & --- & 2.55 & --- & 2.35 & --- \\
$h$ \hfill [$10^{-3}$]      & 3.26 & --- & 3.96 & --- & 3.45 & --- & 3.03 & --- & 2.77 & --- \\
$n_{s}$ \hfill [$10^{-3}$]  & 7.18 & --- & 6.89 & --- & 5.04 & --- & 5.07 & --- & 3.74 & --- \\
$\ombh$ \hfill[$10^{-5}$]   & 12.7 & --- & 15.6 & --- & 11.8 & --- & 9.7 & --- & 7.4 & --- \\
$\omch$\hfill [$10^{-4}$]   & 8.81 & --- & 10.14 & --- & 8.82 & --- & 8.1 & --- & 7.35 & --- \\
$\taure$ \hfill [$10^{-3}$] & 6.85 & --- & 6.79 & --- & 6.57 & --- & 6.64 & --- & 6.27 & --- \\
\enddata
\end{deluxetable*}
}
%%%%%%%%%%%%%%%%%%%%%%%%%%%%%%%%%%%%%%%%%%%%%%%%%%%%%%%%%%%%%%%%%%%%%%%%%%%%%%%%
%%%%%%%%%%%%%%%%%%%%%%%%%%%%%%%%%%%%%%%%%%%%%%%%%%%%%%%%%%%%%%%%%%%%%%%%%%%%%%%%
%%%%%%%%%%%%%%%%%%%%%%%%%%%%%%%%%%%%%%%%%%%%%%%%%%%%%%%%%%%%%%%%%%%%%%%%%%%%%%%%
%LCDM+single-parameter ext
\newcommand{\lcdmplussingleparamext}{
\begin{deluxetable*}{lccccc}
\tabletypesize{\small}
%\tablewidth{0.5\textwidth}
%\def\arraystretch{1.1}
%\renewcommand{\arraystretch}{1.}
\tablecaption{Forecasted $1\sigma$ errors on single-parameter extensions to the \llcdm\ for various SPT-3G surveys in combination with \planck{}.}
\label{tab_ext_param_table}
\tablehead{
Parameter & \mainfield{} & \summerfield{} & \extfield{} & \mainplussummerfield{} & \allspt{}
}
\startdata
\hline \hline
$\neff$  \hfill [$10^{-2}$] & 10.1  & 10.9  & 9.23  & 8.27 & 6.75 \\
$\nrun$ \hfill [$10^{-3}$] & 4.94  & 4.99  & 4.59  & 4.44 & 3.93 \\
$\omk$\hfill [$10^{-3}$] & 2.46  & 2.49  & 2.38  & 2.35 & 2.26 \\
$\summnu$ [eV] \hfill[$10^{-2}$] & 5.99  & 6.15  & 5.91  & 5.78 & 5.57 \\
$\wde$ \hfill [$10^-2$] & 10.23 & 10.44 & 10.03 & 9.91 & 9.64 \\
$\yp$ \hfill [$10^{-3}$] & 6.48  & 6.68  & 5.75  & 5.34 & 4.43 \\
\enddata
\end{deluxetable*}
}
%%%%%%%%%%%%%%%%%%%%%%%%%%%%%%%%%%%%%%%%%%%%%%%%%%%%%%%%%%%%%%%%%%%%%%%%%%%%%%%%
%%%%%%%%%%%%%%%%%%%%%%%%%%%%%%%%%%%%%%%%%%%%%%%%%%%%%%%%%%%%%%%%%%%%%%%%%%%%%%%%
%%%%%%%%%%%%%%%%%%%%%%%%%%%%%%%%%%%%%%%%%%%%%%%%%%%%%%%%%%%%%%%%%%%%%%%%%%%%%%%%

%LCDM degradation with single parameter extensions
\newcommand{\lcdmtabledegradationwithextensions}{
\begin{deluxetable*}{lcccccc}
\tabletypesize{\small}
%\tablewidth{2\textwidth}
%\def\arraystretch{1.2}
%\renewcommand{\arraystretch}{1.}
\tablecaption{Degradation in constraints on standard cosmological parameters when the \llcdm\ model is extended by single additional parameters. Note that $H_0$ is poorly constrained when $\wde$ is allowed to vary.}
\label{tab_lcmd_degradation_due_to_ext}
\tablehead{
Parameter & $A_{s}$[$10^{-11}$] & $H_0$ [$10^{-1}$] & $n_{s}$ [$10^{-3}$]   & $\ombh$[$10^{-5}$] & $\omch$ [$10^{-4}$]  & $\taure$[$10^{-3}$]   
}
\startdata
\hline\hline
\llcdm\ &  2.29  &  2.78  &  2.83  &  5.67  &  7.10  &  6.19 \\
$\neff$                                      &   2.51   &   6.18     &   4.88   & 9.61    &   10.80       &   6.39       \\
$\nrun \equiv {\rm d}n_s/{\rm d}{\rm ln} k$  &   2.43   &   2.83     &   2.94   & 6.90    &   7.14        &   6.35       \\
$\omk$                                       &   2.55   &   8.73     &   3.10   & 5.73    &   9.56        &   6.42       \\
$\summnu$[eV]                                &   2.78   &   7.84     &   3.00   & 5.77    &   9.19        &   6.90       \\
$\wde$                                       &   2.75   &   28.35    &   2.72   & 5.62    &   6.53        &   7.02        \\
$\yp$                                        &   2.52   &   3.31     &   4.44   & 9.34    &   7.49        &   6.67       \\
\enddata
\end{deluxetable*}
}
%%%%%%%%%%%%%%%%%%%%%%%%%%%%%%%%%%%%%%%%%%%%%%%%%%%%%%%%%%%%%%%%%%%%%%%%%%%%%%%%
%%%%%%%%%%%%%%%%%%%%%%%%%%%%%%%%%%%%%%%%%%%%%%%%%%%%%%%%%%%%%%%%%%%%%%%%%%%%%%%%
%%%%%%%%%%%%%%%%%%%%%%%%%%%%%%%%%%%%%%%%%%%%%%%%%%%%%%%%%%%%%%%%%%%%%%%%%%%%%%%%

%\title{Forecasts of Constraints on Cosmological Parameters from using Primary and Lensed Cosmic Microwave Background Signals in \sptnew{} Maps}
%\title{Constraints on Cosmological Parameters expected from gravitationally-lensed primary Cosmic Microwave Background signals in SPT-3G maps}
%\title{Tests of $\mathbf{\Lambda}$CDM and Extensions Expected from the Gravitationally-Lensed Primary Cosmic Microwave Background Signals in SPT-3G Maps.}
%\title{Tests of $\mathbf{\Lambda}$CDM to be enabled by SPT-3G observations of the gravitationally-lensed cosmic microwave background}
%\title{Testing $\mathbf{\Lambda}$CDM with forthcoming SPT-3G measurements of CMB and CMB lensing power spectra}
\title{Testing the $\mathbf{\Lambda}$CDM Cosmological Model with Forthcoming Measurements of the Cosmic Microwave Background with SPT-3G}
\author{K.~Prabhu}
\affiliation{Department of Physics \& Astronomy, University of California, One Shields Avenue, Davis, CA 95616, USA}

\author[0000-0003-1405-378X]{S.~Raghunathan}
\affiliation{Center for AstroPhysical Surveys, National Center for Supercomputing Applications, Urbana, IL, 61801, USA}

\author[0000-0001-7317-0551]{M.~Millea}
\affiliation{Department of Physics, University of California, Berkeley, CA, 94720, USA}

\author{G.~P.~Lynch}
\affiliation{Department of Physics \& Astronomy, University of California, One Shields Avenue, Davis, CA 95616, USA}
\author{P.~A.~R.~Ade}
\affiliation{School of Physics and Astronomy, Cardiff University, Cardiff CF24 3YB, United Kingdom}

\author{E.~Anderes}
\affiliation{Department of Statistics, University of California, One Shields Avenue, Davis, CA 95616, USA}

\author[0000-0002-4435-4623]{A.~J.~Anderson}
\affiliation{Fermi National Accelerator Laboratory, MS209, P.O. Box 500, Batavia, IL, 60510, USA}
\affiliation{Kavli Institute for Cosmological Physics, University of Chicago, 5640 South Ellis Avenue, Chicago, IL, 60637, USA}
\affiliation{Department of Astronomy and Astrophysics, University of Chicago, 5640 South Ellis Avenue, Chicago, IL, 60637, USA}

\author{B.~Ansarinejad}
\affiliation{School of Physics, University of Melbourne, Parkville, VIC 3010, Australia}

\author[0000-0002-0517-9842]{M.~Archipley}
\affiliation{Department of Astronomy, University of Illinois Urbana-Champaign, 1002 West Green Street, Urbana, IL, 61801, USA}
\affiliation{Center for AstroPhysical Surveys, National Center for Supercomputing Applications, Urbana, IL, 61801, USA}

\author[0000-0001-6899-1873]{L.~Balkenhol}
\affiliation{Sorbonne Universit\'e, CNRS, UMR 7095, Institut d’Astrophysique de
Paris, 98 bis bd Arago, 75014 Paris, France}

% ...

\begin{abstract}
We forecast constraints on cosmological parameters enabled by three surveys conducted with \sptnew{}, the third-generation camera on the South Pole Telescope. The surveys cover separate regions of \mbox{1500}, \mbox{2650}, and \mbox{6000 $\sqdeg$} to different depths, in total observing 25\% of the sky.  These regions will be measured to white noise levels of roughly $2.5$, $9$,  and $12\ \ukarcmin$, respectively, in CMB temperature units at 150\,GHz by the end of 2024.  The survey also includes measurements at 95 and 220\,GHz, which have noise levels a factor of $\sim$1.2 and 3.5 times higher than 150\,GHz, respectively, with each band having a polarization noise level $\sim \sqrt{2}$ times higher than the temperature noise. We use a novel approach to obtain the covariance matrices for jointly and optimally estimated gravitational lensing potential bandpowers and unlensed CMB temperature and polarization bandpowers. We demonstrate the ability to test the \llcdm\ model via the consistency of cosmological parameters constrained independently from \sptnew{} and \planck\ data, and consider the improvement in constraints on \llcdm\ extension parameters from a joint analysis of \sptnew{} and \planck\ data. The \llcdm\ cosmological parameters are typically constrained with uncertainties up to $\sim$2 times smaller with \sptnew\ data, compared to \planck, with the two data sets measuring  significantly different angular scales and polarization levels, providing additional tests of the standard cosmological model.
\end{abstract}
%\keywords{cosmic background radiation - cosmological parameters - Forecasting}

%--------------------------------------------------------------------------%

%\section*{To Do}
%\begin{enumerate}

%\item {\pending{Lennart's suggestions:} Figs: 8 and 9 modifications; Fig. 12: Make it two panels and add SPT-3G 2018/Planck to show that all models are consistent with them. Work in progress.}
%\item {\pending{Lennart's suggestions:} Test the importance of Tcal;  correlated 1/f noise for summer/wide fields; Add a footnote to say why $\ell-\ell^{\prime}$ is not that important (Lloyd); Make sure to talk about lensing degeneracy in the main text.}
%\end{enumerate}

\section{Introduction} \label{sec_intro}
    Observations of the cosmic microwave background (CMB) temperature and polarization anisotropies have played a crucial role in the field of cosmology, particularly in the establishment of a six-parameter standard cosmological model, \llcdm. This model which provides an excellent fit to CMB data from \planck{} \citep{planck18-6}, the Atacama Cosmology Telescope (ACT) \citep{aiola20, choi20, madhavacheril23} and the South Pole Telescope (SPT) \citep[][]{simard18, dutcher21, balkenhol21, pan23}, as well as a great variety of other astrophysical datasets
    %such as
    \citep[][]{alam21, descollaboration24, abbott22a, kuijken19}, has been enormously successful in providing a consistent framework for understanding the universe. 
     %At the same time, this empirical success is a mystery, as critical ingredients (dark matter, dark energy, and the generator of primordial fluctuations) lie beyond the standard model of particle physics. 
     
    These successes are perhaps surprising, since critical ingredients to the model (dark matter, dark energy, and the generator of primordial fluctuations) appear disconnected from the physics we know of through laboratory experiments. The desire for clues that could deepen our understanding of these ingredients motivates us to continue our testing of the model. One particularly compelling line of testing emerges from the fact that the standard cosmological model, conditioned on data from the \planck{} satellite, makes extremely precise predictions for the CMB temperature, polarization, and lensing power spectra on angular scales not well measured by \planck{}. 

   Additional motivation for testing \llcdm{} (beyond our ignorance about dark matter, dark energy, and the generator of primordial fluctuations) comes from cosmological tensions and anomalies as these might be evidence for beyond-\llcdm\ physics.  A prominent issue is the Hubble tension – a notable discrepancy between classical distance ladder determinations \citep{murakami23,riess22} and \llcdm-dependent determinations  \citep{planck18-6} of the current expansion rate of the universe. 
    An additional, though statistically weaker tension, is in the values of $\sigma_8$, the root mean square of the density field smoothed over \mbox{$8h^{-1}$ Mpc} radius spheres, derived from weak lensing observations in optical galaxy surveys and those inferred from CMB data\footnote{Although this tension is partially resolved in the analysis of \planck{} PR4 maps \citep{tristram24}} \citep{asgari21, joudaki20, abbott22a}. 
    There are also peculiar patterns in the CMB temperature anisotropies on large angular scales, which, by some calculations, are extremely unlikely in a \llcdm\ universe \citep{copi10, givans23}.
    %Therefore, experiments such as \sptnew{}, ACT, and \planck{} which measure relatively independent signals serve as crucial cross-validation tools and allow for an in-depth investigation into these tensions. 

    Altogether these discrepancies offer some evidence that \llcdm\ is not the final word in cosmology, encouraging us to search for beyond-\llcdm\ signals in our surveys. Indeed, 
    there are alternative models that reduce the $\sigma_{8}$ or $H_0$ tensions and make different predictions for the power spectra than is the case for \llcdm. In this context, it becomes pertinent to investigate the extent to which upcoming observational data can distinguish between the standard \llcdm\ model and some of these alternatives.
    %We qualitatively investigate the potential of the \sptnew{} surveys to distinguish between the \llcdm\ model and some of these alternatives.

    %paragraph entirely focused on testing LCDM
    %When conditioned on prior data, primarily from the \planck{} satellite, the standard cosmological model makes extremely precise predictions for the statistical properties of signals not well measured by \planck{} in polarization anisotropies and, on small scales, in temperature anisotropies. These precise predictions offer the opportunity for new measurements to make powerful tests of the \llcdm\ model. 

    %In this paper we study how the full data set from \sptnew{}, the third-generation camera on the SPT, will realize this potential.
    \sptnew, the third-generation camera on the SPT, has the potential to test the $\lcdm$ model by precisely mapping the primary and lensing anisotropies of the CMB.
    The full \sptnew{} data set will comprise three surveys, which, altogether, cover 10000\,$\sqdeg$, or $\sim$25\% of the sky. The \mainfield\ survey (1500\,$\sqdeg$), observed over five austral winters, is approaching the target depths of the next generation CMB-S4 Deep and Wide survey of 70\% of the sky planned to begin next decade \citep{cmbs4collab22}. Next-deepest is the \summerfield{} survey, a completed collection of three fields observed during four austral summers totaling 2650\,$\sqdeg$.
    The shallowest is the 6000\,$\sqdeg{}$ \extfield{} survey, which, after one observing season, will still be deeper than surveys from any other existing high-resolution CMB experiment of comparable survey size. 

    %secondaries ++
    Besides mapping the primary CMB and lensing anisotropies, \sptnew{} will provide a powerful set of data for studying a broad range of topics in cosmology and astrophysics. 
    These include the delensing of lensing-induced B-modes \citep{bicep2keck21}; production of mass-limited catalogs of galaxy clusters out to high redshifts using the thermal Sunyaev-Zel{'}dovich (tSZ) signature \citep{bleem15b, huang20, bleem20, bleem23}, and constraining cosmology with these catalogs \citep{bocquet24}; robust measurements of the kinematic Sunyaev-Zel{'}dovich (kSZ) signal for velocity reconstruction and constraining the epoch of reionization \citep{schiappucci23, raghunathan23, raghunathan24}; production of mm-wave point source catalogues \citep{everett20}; and the
    detection and monitoring of mm-wave galactic, extragalactic, and solar system transient, variable, and moving objects \citep{whitehorn16, guns21, chichura22, hood23, tandoi24}. 
    
    %We demonstrate the capacity of these surveys to test \llcdm\ in several ways. 
    Here we demonstrate the capacity of the \sptnew{} surveys to test \llcdm\ in several ways, focusing exclusively on the constraining power of CMB and lensing power spectrum measurements. 
    We forecast how well we will measure the temperature and $E$-mode polarization power spectra, their cross power spectrum, and the power spectrum of the lensing potential. %We will see that these measurements are expected to be significantly tighter than existing measurements on angular scales where the predictions of the \llcdm\ model are highly precise, and also where the error bars are smaller than the differences between these predictions and those of some alternative models. 
    
    We also propagate the power spectrum uncertainties forward to constraints on cosmological parameters. 
    From \sptnew{} data alone we can obtain constraints on the parameters of the \llcdm\ model that are comparable to, and in some cases, better than those from \planck. 
    When compared to \planck{}, a greater fraction of the weight of these constraints comes from polarization \citep[see][for example]{galli14}, from  smaller angular scales, and from lensing. 
    Therefore, comparison of these parameter estimates with those from \planck{} will be an excellent test of the \llcdm\ model. 
    
    The outcome of such a test could provide evidence for physics beyond \llcdm.
    Looking ahead to this scenario, we also forecast constraints on single- and double-parameter extensions of \llcdm\ from the combination of \planck{} and \sptnew{} data and report how these constraints improve upon those from \planck{} alone. 
    The test could equally well reveal consistency with \llcdm, an outcome that should not be dismissed given the model's track record.
    If this is the case, joint \planck{}+\sptnew{} \llcdm\ parameter constraints will be of interest, and we provide forecasts for the combined constraints on \llcdm\ as well.
    % and report how these constraints improve upon those we have from \planck{} alone. 
    
    %These tests, when we have the data to conduct them and have completed their reduction, may reveal consistency with \llcdm, an outcome that should not be dismissed given the model's track record. If this is the case, joint \planck{} + SPT-3G \llcdm\ parameter constraints will be of interest. %We therefore also provide forecasts for their errors.  

     %Additionally, there are observed inconsistencies in the preferred cosmological models derived from CMB data at large and small angular scales \citep{planck18-6, Aylor_2019, Addison_2018}. 
    In addition to our comparison with \planck, we compare with forecasts we make for the high-resolution survey of the Simons Observatory (SO, \citealt{simonsobservatorycollab19}). SO is expected to survey roughly 40\% of the sky (much more sky is available to survey from mid-latitude sites such as the SO site in the Atacama Plateau in Chile compared to the South Pole) to noise levels similar to or slightly better than the SPT-3G Summer fields, leading to constraints on \llcdm\ parameters that are comparable to what we expect from the combined \sptnew{} surveys.
    
    %In addition to \planck, we also compare the \sptnew\ constraints to our forecast for the high-resolution survey for the Simons Observatory (SO, \citealt{simonsobservatorycollab19}), scheduled to begin in 2024 at the earliest.
    %In addition to comparing with \planck\ we compare with forecasts we make for the high-resolution survey of the Simons Observatory (SO, \citealt{simonsobservatorycollab19}), scheduled to begin in 2024. 
    %From its mid-latitude site in the Atacama plateau of Chile, the SO can access much more sky than is possible from the South Pole. 
    %We find that the benefit of SO's larger sky coverage largely offsets the impact of its higher map noise levels, leading to constraints on \llcdm\ parameters that are comparable to what we expect from the combined \sptnew{} surveys. 

%%%%%%%%%%%%%%%%%%%%%%%%%%%%%%%%%%%%%%%%%%%%%%%%%%%%%%%%%%%%%%%
\commenter{
  \kp{ Revise this: We also compare with the projected uncertainty from SO, finding that \sptnew{} will deliver constraints that are slightly looser than those to come from SO on multipoles of 300 or smaller, and constraints that are significantly tighter than those to come from SO on higher multipoles.}

This increased sensitivity to CMB lensing is due to the depth of these surveys, and the \mainfield{} survey in particular. Gravitational lensing deflections distort the statistical properties of the observed temperature and polarization maps, maps that would otherwise be expected to be Gaussian and statistically isotropic random fields. These distortions have been used to infer deflection maps and their power spectra \citep{smith07, vanengelen12, omori17, wu19, madhavacheril23, qu24, pan23, das11}
Such power spectra bring an additional sensitivity to late-time growth of structure and distance-redshift relations. The deflection maps themselves can be used in cross correlation with other surveys to probe the relationship between mass and light \citep{bleem12b, sherwin12, holder13, omori23}. 
    
    %Delensing
The process of delensing, reconstructing the temperature and polarization maps as they would be in the absence of gravitational lensing, also has several advantages. These include tighter constraints on parameters that impact the damping tail, simplified analyses due to reduced bandpower covariances, tighter constraints on non-Gaussianity, and reduced sensitivity to non-linear and baryonic feedback effects \citep{meyers16, hotinli22}. Delensed maps have been made from \planck{} data \citep{carron17, han2020}, although the noise levels are too high for any of these advantages to be significant. 

Delensing will soon make a difference for constraints on primordial gravitational waves (PGWs) from polarization $B$ modes. 
\citet{manzotti17} demonstrated the first reduction in $B$-mode power to come from removing the $B$ modes generated from $E$ modes by gravitational lensing \citep[Also see][]{bicepkeckspt21}. 
The tightest published constraints to date are from the BICEP/{\it Keck} (BK) Collaboration \citep{bicep2keck21}, include data through their 2018 observing season, and are only mildly degraded by the additional ``noise" of $B$ modes generated from $E$ modes by gravitational lensing.
Data collected by the BK Collaboration since 2018 has been used to make maps with noise levels below the level of the lensing $B$-mode noise. Hence, we expect that the next limits they publish on PGWs will either be degraded by lensing $B$ modes, or significantly improved by de-lensing with \sptnew{} data \citep{bicepkeckspt21}. 

    %MUSE
    For an ongoing analysis of the first two full years of data from the \mainfield{} survey, one of our analysis approaches is a Bayesian one, which, in principle, is unbiased and optimal (information lossless), despite the presence of gravitational lensing, and naturally results in joint constraints on the lensing power spectrum and unlensed CMB power spectra. Central to the approach, is our construction of a likelihood model, in which the unlensed CMB fields (in temperature and polarization) and a deflection field, are statistically isotropic and Gaussian. Comparison with the data in the likelihood is done with forward modeling of how the unlensed CMB fields, the lensing field, and various instrumental effects, combine to produce the signal in our maps. To speed up estimation of the unlensed and lensing power spectra from this likelihood we employ the Marginal Unbiased Score Expansion (MUSE) technique \citep{millea21_muse}. \lloyd{We refer to this whole inference pipeline from maps to the bandpowers and their covariance matrix as MUSE.}

    Since MUSE produces a bandpower covariance matrix for all the relevant spectra, it is quite convenient for us to propagate these uncertainties to a parameter error covariance matrix, and compare the results with more traditional forecasting techniques. We find, consistent with expectations based on other work \citep{meyers16}, that any gains for \sptnew{} are not dramatic; i.e., the traditional forecasting and MUSE-based forecasting lead to errors that agree at the 10\% level. We will discuss the situations in which we can expect Bayesian (or otherwise near-optimal) approaches to make a significant difference, as well as other advantages of a MUSE-based parameter forecast.
    }
%%%%%%%%%%%%%%%%%%%%%%%%%%%%%%%%%%%%%%%%%

%MUSE
For an ongoing analysis of the first two full years of data from the \mainfield{} survey, one of our analysis approaches is a Bayesian one, which is in principle unbiased and optimal (information lossless) despite the presence of gravitational lensing, and naturally results in joint constraints on the lensing power spectrum and unlensed CMB power spectra. 
To speed up estimation of the unlensed and lensing power spectra, from this likelihood, we employ the Marginal Unbiased Score Expansion (MUSE) technique \citep{millea21_muse}. 
We refer to this whole inference pipeline from maps to the bandpowers and their covariance matrix as MUSE.
Since MUSE can be used to obtain a covariance matrix for all the relevant spectra, it is quite convenient for us to propagate these uncertainties to a parameter error covariance matrix, and compare the results with more traditional forecasting techniques. 

    This paper is structured as follows: In Sec.~\ref{sec_exp_setups}, we describe the specifications of the \sptnew{} instrument, the survey regions and our assumptions about the noise and foregrounds. In Sec. \ref{sec_methodology} we describe the methodology used to obtain these forecasts. We report the forecasts of the parameter constraints for \llcdm\ with \sptnew{} alone and for extensions with \sptnew{} combined with \planck{} in Sec. \ref{sec_results}. In Sec. \ref{sec_lcdm_alternatives}, we discuss prospects for \sptnew{} to detect departures from \llcdm\ predictions and test alternate models. Finally, we summarize our findings in Sec. \ref{sec_conclusion}.

\refresponse{
In this work, we set the fiducial values of the baseline $\lcdm$ cosmology and its extensions to \planck{} 2018 measurements (TT, TE, EE + lowE + lensing in Table 2 of \citealt{planck18-6}) and compute the theoretical CMB power spectra using \texttt{CAMB} software \citep{lewis00}.
}
%--------------------------------------------------------------------------%

\section{Experiment Specifications}
\label{sec_exp_setups}

\begin{figure*}
\centering
\ifdefined\apjformat
    \includegraphics[width=0.8\textwidth]{survey_footprints.pdf}
\else
    \includegraphics[width=0.8\textwidth]{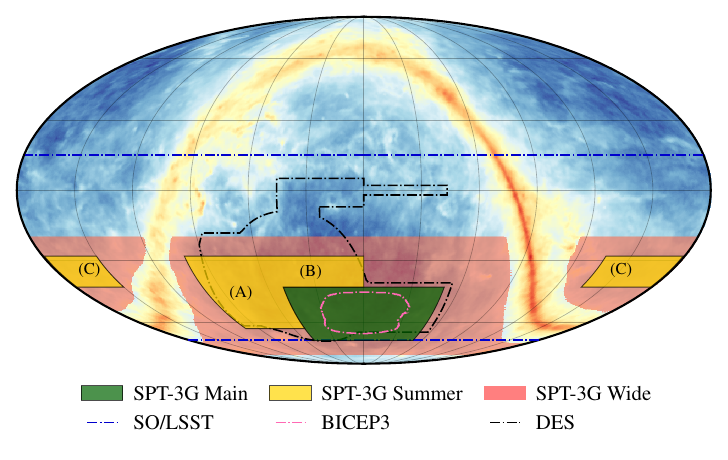}
\fi
\caption{
Footprints of the three \sptnew{} surveys: \mainfield{} (green), \summerfield{} (yellow), and \extfield{} (red). 
%Note that the \mainplussummerfield{} is the combination of the \mainfield{} (5Y) and the \summerfield{}(4Y) surveys. 
%\allspt{} survey is a combination of all the three \sptnew{} surveys.
Also shown are the footprints of other surveys: BICEP3 (pink dash-dotted), DES (black dash-dotted), and SO/LSST (blue dash-dotted). The background is the map of the galactic dust produced by the \planck{} satellite \citep{planck16_comp_sep}.
}
\label{fig_spt_footprints}
\end{figure*}

In this section, we briefly describe the specifications of the \sptnew{} experiment along with the details of the three survey regions and a description of the noise characteristics.

\subsection{Survey Specifications}
\label{subsec_survey_specs}
\begin{deluxetable*}{l | c | c | c| c| cccc}
\tabletypesize{\small}
%\tablewidth{0.5\textwidth}
%\def\arraystretch{1.2}
%\renewcommand{\arraystretch}{1.}
\label{tab_survey_specs}
\tablecaption{Sky fraction and temperature white noise levels in the three bands for different SPT surveys considered in this work. The polarization white noise levels are expected to be $\sim$$\sqrt{2}$ times higher. }
\tablehead{
     \multirow{2}{*}{Survey} & Area & RA & Dec. & \multirow{2}{*}{Years observed} & \multicolumn{4}{c}{Noise level ($\Delta_{T}$) [\ukarcmin]}  \\
     \cline{6-9}
     & [$\sqdeg$] & center [deg] & center [deg] & & 95\,GHz & 150\,GHz & 220\,GHz & Coadded
}
\startdata
     \hline\hline
     \multicolumn{7}{l}{{\it Completed:}}\\\hline
     \sptnew\ Main & 1500 & 0 & -57.5 & \multirow{4}{*}{2019-2023} & 3.0 & 2.5 & 8.9 & 1.9 \\
     %\hline
       %Summer & 2640 & \multicolumn{2}{c|}{-} & &  &  &  &  \\
      \sptnew{} Summer-A & 1210 & 75.0 & -42.0 & & 9.4 & 8.7 & 29.5 & 6.2 \\
      \sptnew{} Summer-B & 570 & 25 & -36 & & 10.0 & 9.7 & 30.3 & 6.8 \\
      \sptnew{} Summer-C & 860 & 187.5 & -38 & & 10.0 & 9.2 & 27.8 & 6.6 \\\hline\hline
     \multicolumn{7}{l}{{\it Ongoing+Future:}}\\\hline
      \multirow{2}{*}{\sptnew{} Main}  & \multirow{2}{*}{1500} & \multirow{2}{*}{0} & \multirow{2}{*}{-57.5} & 2019-2023 & \multirow{2}{*}{2.5} & \multirow{2}{*}{2.1} & \multirow{2}{*}{7.6} & \multirow{2}{*}{1.6} \\
       &  & & & + 2025-2026 & & & & \\
      \cline{3-4}
      \sptnew\ Wide & 6000 & \multicolumn{2}{c|}{Multiple} & 2024 & 14 & 12 & 42 & 8.8
\enddata
\end{deluxetable*}

\begin{deluxetable}{c|c|c|c}
%\tabletypesize{\small}
%\tablewidth{0.5\textwidth}
%\def\arraystretch{1.2}
%\renewcommand{\arraystretch}{1.}
\tablecaption{Atmospheric noise specifications for the three SPT bands.}
\label{tab_beam_atmnoise_specs}
\tablehead{
        \multirow{2}{*}{Specification} & \multicolumn{3}{c}{Observation band}\\
        \cline{2-4}
        & 90\,GHz & 150\,GHz & 220\,GHz
}
\startdata
\hline\hline
        \multicolumn{4}{l}{SPT-3G Main:}\\\hline
        {$\ell_{\text{knee},T}$} &  1200 & 2200 & 2300 \\\hline
        $\alpha_T$ & -3 & -4 & -4 \\\hline
        $\ell_{\text{knee},P}$ & \multicolumn{3}{c}{300}\\\hline
        $\alpha_P$ & \multicolumn{3}{c}{-1}\\\hline\hline
        \multicolumn{4}{l}{SPT-3G Summer \& SPT-3G Wide:}\\\hline
        {$\ell_{\text{knee},T}$} &  1600 & 2600 & 2600 \\\hline
        $\alpha_T$ & -4.5 & -4 & -3.9 \\\hline
        $\ell_{\text{knee},P}$ & 300 & 490 & 500 \\\hline
        $\alpha_P$ & -2.2 & -2 & -2.5
\enddata
\end{deluxetable}

\begin{figure*}
\centering
\ifdefined\apjformat
    \includegraphics[width=0.8\textwidth, keepaspectratio]{spt_noise_ilc_curves.pdf}

\else
    \includegraphics[width=0.8\textwidth, keepaspectratio]{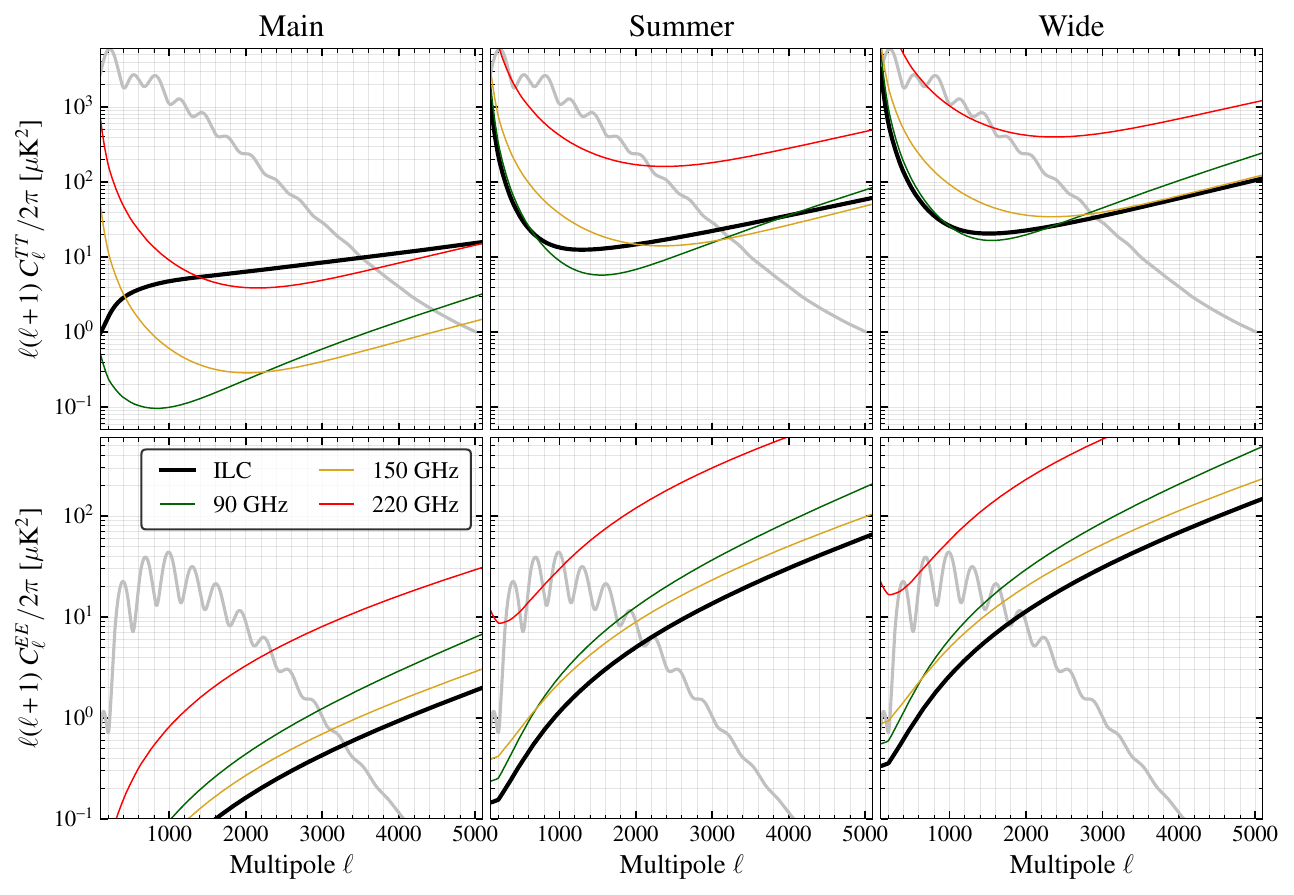}
\fi
\caption{Expected $TT$ and $EE$ noise power spectra for the three SPT-3G surveys: Beam-deconvolved noise curves in the three bands for SPT-3G Main (5 years), Summer (4 years) and Wide (1 year) surveys are shown as colored curves. The Minimum Variance (MV) combination of the LC residuals is the thick black solid curve in all the panels. The gray curves represent the CMB power spectra. Top panels are for temperature and the bottom panels are for polarization. The LC residuals are dominated by foregrounds for temperature for \mainfield{} while for polarization they are roughly equal to the inverse variance weighted noise spectra from the three bands.}
\label{fig_noise_ilc_residuals}
\end{figure*}

\begin{figure}
\ifdefined\apjformat
    \includegraphics[width=0.45\textwidth, keepaspectratio]{snr_spt_planck.pdf}
\else
    \includegraphics[width=0.45\textwidth, keepaspectratio]{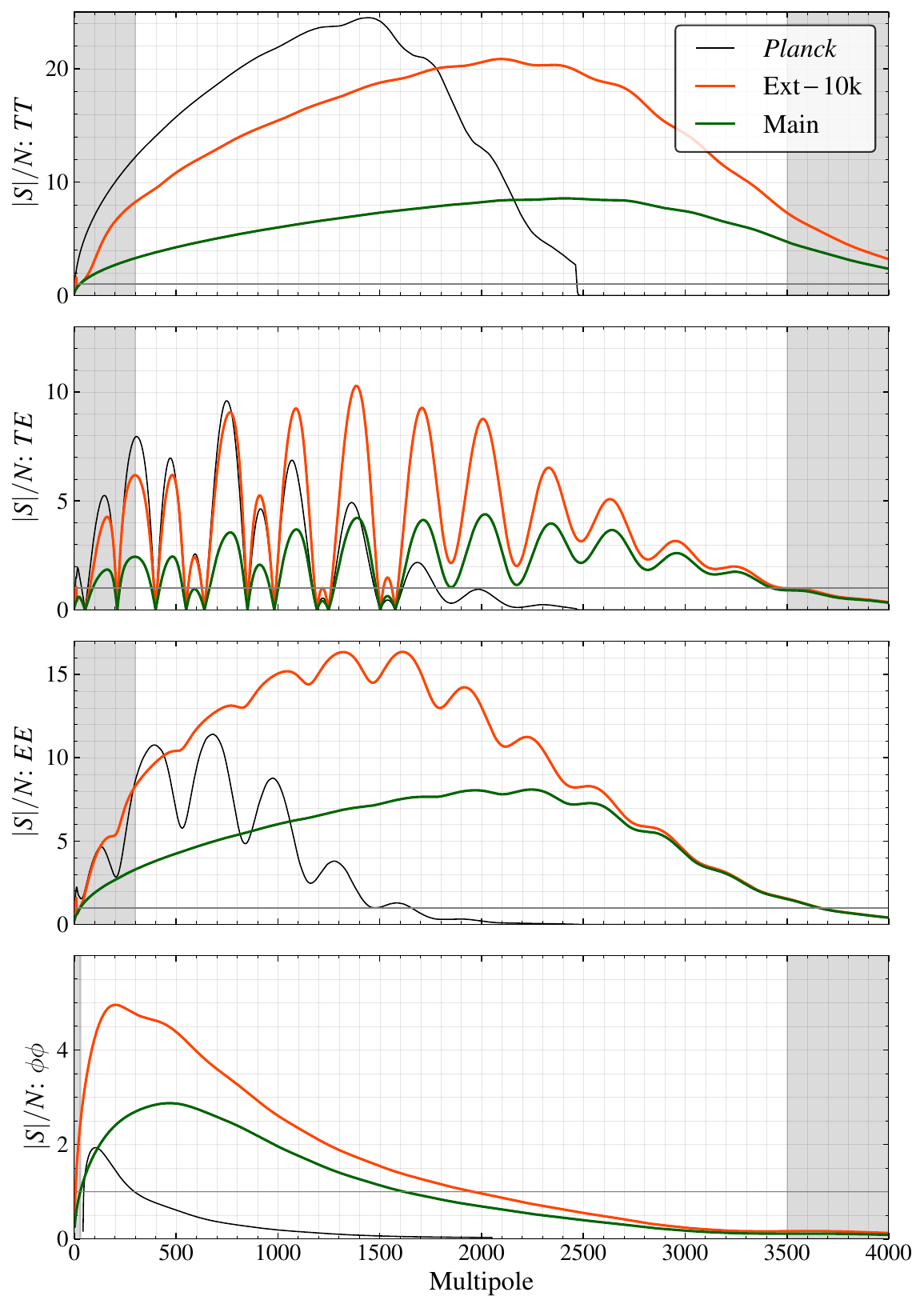}
\fi
\caption{$\snr$ per multipole on the primary and lensing CMB power spectra for the \mainfield{} (green) and the \allspt{} (orange) \sptnew\ surveys, with values for \planck{} (black) plotted for comparison. 
The constraining power of the full \allspt{} survey will be larger than \planck{} at $\ell \gtrsim 1800$ in $TT$, $\ell \gtrsim 800$ in $TE$, $\ell \gtrsim 450$ in $EE$ and effectively for all the modes in $\phi \phi$. The non-shaded area corresponds to the range of multipoles used in the forecasts.
}
\label{fig_powspec_sn}
\end{figure}

%We present a brief description about the South Pole Telescope (SPT) and the \sptnew{} camera here. 
SPT is a 10-meter diameter telescope specifically engineered for low-noise and high-resolution measurements of the millimeter-wave sky. 
%Situated at the Amundsen-Scott South Pole Station, the telescope takes advantage of the site's low atmospheric fluctuation power on degree angular scales, making it one of the best locations on Earth for millimeter-wave observations. 
The telescope is currently equipped with \sptnew, the third CMB camera installed on SPT. \sptnew\ is a significant upgrade compared to previous cameras, incorporating a polarization sensitive tri-chroic focal plane with almost 16,000 detectors, operating in three frequency bands, 95, 150, and 220 GHz.
For more details about the instrument, we refer the readers to \cite{benson14, bender18, anderson18, sobrin18, sobrin22}.

%\sptnew{} has been surveying a 1500 $\sqdeg${} field (\mainfield) since 2019 during the austral winter. 
%%%%%%%%%%%%%%%%%%%%%%%%%%%%%%%%%%%%%%%
%Tom suggested to replace these paragraphs
\commenter{
\sptnew{} has been fully operational since 2019 surveying a 1500 $\sqdeg${} field (\mainfield) during the austral winters. 
During austral summer, the Sun enters our \mainfield{} field, so we switch the austral summer observations to a 2650 $\sqdeg${} field (\summerfield) every year. \summerfield{} survey consists of three sub-fields, namely, Summer-a, Summer-b, and Summer-c. The \mainfield{} survey completely overlaps with the BK patch. 
It also has a $\sim 75\%$ overlap with Dark Energy Survey (DES). 
The \summerfield{} survey, on the other hand, has a much bigger overlap with the DES. 

The observational synergies with other experiments have enabled the current and past SPT surveys to pursue a variety of science goals besides the standard CMB and lensing power spectra measurements. 
In fact, one of the main goals of \sptnew{} is to construct a deep CMB lensing map, which can be used to the delens the lensing-induced B-modes and facilitate measurements of the primordial B-modes by BK. 
While this requires deep CMB polarization maps like those from \sptnew \mainfield{} survey, improving the precision on many of the cosmological parameters demands observations of larger sky patches, as we demonstrate in this work. 
To facilitate this, \sptnew{} is currently scanning an additional 6000 $\sqdeg${} field (\extfield).
}
%%%%%%%%%%%%%%%%%%%%%%%%%%%%%%%%%%%%
One of the primary science goals of SPT-3G is to create a template of the gravitational-lensing-induced B modes in the patch of sky observed by the BICEP-Keck (BK) family of experiments. This template can be subtracted from the BK data, reducing the primary source of variance in the BK B-mode analysis and potentially improving the sensitivity to primordial gravitational waves by a factor of three. 
To that end, since 2019, SPT-3G has been used primarily (eight months per year) to observe the 1500 $\sqdeg${} \mainfield{} survey, which overlaps completely with the BICEP3 and BICEP Array sky coverage (see Fig.~\ref{fig_spt_footprints}). 
During the austral summer season (roughly at the beginning of December), when the Sun is above the horizon and gets sufficiently close to the \mainfield{} survey region, it is picked up by the telescope far sidelobes . Therefore, until sunset, we switch to observing three fields that are safe from sun contamination. These comprise the 2650 $\sqdeg${} \summerfield{} survey (Fig.~\ref{fig_spt_footprints}).
We refer to these three fields as Summer-A, Summer-B, and Summer-C. 

Even with the added sky coverage of the Summer survey, the constraints on \llcdm\ parameters from \sptnew{} data are limited by sample variance. For this reason, we have paused our standard observing strategy for one year to undertake a new survey of all the sky viewable from the South Pole at instrumentally feasible observing elevations and with minimal Galactic foreground contamination. We have empirically determined that, during the winter, we are able to observe with \sptnew{} between elevations of roughly $20^\circ$ and $80^\circ$ (limited by detector linearity at the low end and refrigerator performance on the high end). The intersection of these elevation limits (which, at the South Pole, correspond directly to declination limits) and the \planck\ \texttt{GAL080} \citep{planck15-13} Galactic mask defines our new \extfield{} survey, which totals 6000 $\sqdeg${} and is shown in Fig.~\ref{fig_spt_footprints}. 
Observations of the \extfield{} survey commenced in December of 2023 and will continue through the austral winter of 2024. We will refer to the combination of the \mainfield{} and the \summerfield{} surveys as the \mainplussummerfield{} survey. 
Similarly, we will refer to the combination of \mainfield, \summerfield, and \extfield{} surveys as the \allspt{} survey. In Table \ref{tab_survey_specs}, we summarize the areas and depths for the SPT-3G Main, Summer, and Wide surveys.

The survey footprints for \sptnew{} and other surveys are shown in Fig.~\ref{fig_spt_footprints}. 
As is evident from the figure, all the three \sptnew{} surveys have an excellent overlap with SO, which will enable a variety of cross-checks to be performed between the two surveys. 
We list the noise levels and the field centers for the \sptnew{} surveys in Table~\ref{tab_survey_specs}. 
In this work, we produce forecasts both for current noise levels (2019-2023) and for the ones expected from future \sptnew{} observations (2024-2026). 
Besides instrumental noise, our data is also affected by atmospheric noise, which increases towards large angular scales. 
We model the temperature noise power spectrum as
\begin{equation}
 N_\ell =  \Delta_{T}^2 \left[1 + \left(\frac{\ell}{\ell_{\text{knee},T}}\right)^{\alpha_T}\right]\, ,
\label{eq_noise}
\end{equation}
where $\Delta_{T}$ represents the white noise level at a given band in $\ukarcmin$, $\ell_{\text{knee},T}$ and $\alpha_{T}$ are used to parameterize the atmospheric noise. 
For the \mainfield{} survey, we adopt the values of $\ell_{\text{knee}}$ and $\alpha$ in Table \ref{tab_beam_atmnoise_specs} from previous analyses, but we find that the results of our forecasting are insensitive to the exact values of $\ell_{\text{knee}}$ and $\alpha$. For the \summerfield{} survey, during which the atmospheric opacity and precipitable water vapor levels are higher, we fit real noise data to the model in Eq.~\ref{eq_noise}, and the parameters in the Table \ref{tab_beam_atmnoise_specs} are the result of those fits. We also adopt the \summerfield{} survey parameters for the \extfield{} survey (which is likely pessimistic for the higher-elevation Wide fields).
We extrapolate the white noise levels for deeper integration times. 
We also fit the noise model parameters for the polarization maps in a similar way. We note that the white noise levels in the polarization maps are roughly $\sqrt{2}\times$ higher than in the temperature maps.
In Table~\ref{tab_beam_atmnoise_specs}, we list the $\ell_{\text{knee}}$ and $\alpha$ values adopted for different bands for both temperature and polarization.
Finally, we deconvolve the experimental beam window function $B_{\ell}$ from the noise spectra as $N'_{\ell} \equiv \dfrac{N_\ell}{B_{\ell}^2}$ where the beam window function at each frequency is estimated as detailed in \citet{dutcher21}.
%Finally, we deconvolve the experimental beam window function $B_{\ell}$ from the noise spectra as $N'_{\ell} \equiv \dfrac{N_\ell}{B_{\ell}^2}$ where we are using the beam window function calculated from data at each frequency. 
We show the model noise spectra in different bands (90\,GHz in green, 150\,GHz in yellow, and 220\,GHz in red) for all the SPT-3G surveys in Fig.~\ref{fig_noise_ilc_residuals}. 
These spectra assume integration times of 5 years for the \mainfield{} survey, 4 years for the \summerfield survey, and 1 year for the \extfield{} survey.

%--------------------------------------------------------------------------%

\section{Forecasting methodology}
\label{sec_methodology}
In this section, we outline the two methods used for calculating the bandpower covariance matrix from which a Fisher matrix for cosmological parameters \citep{tegmark97a} can be derived. %They both proceed via calculation of a cosmological parameter Fisher matrix from a bandpower covariance matrix. 
The first method relies on an analytic approximation (See \S~\ref{subsec_fisher_formalism}) for the covariance matrix. 
This approach has been widely adopted in various forecasting studies, and is quite simple and flexible. 
Our second method employs MUSE (see \S~\ref{sec_muse}), which is one of the analysis pipelines developed for \sptnew{} data. 
MUSE is more accurate as it readily incorporates non-ideal factors such as covariance between multipoles, covariance between different types of power spectra, and impacts of delensing. Although previous works \citep[for example,][]{hotinli22,trendafilova23}, suggest that these factors may not significantly alter our forecasts, adopting both these methods improves the robustness and reliability of our forecasts.
We also describe the method we use to combine \planck{} data with SPT-3G. We begin with a brief description of how we optimally combine data from multiple frequency bands.

\subsection{Linear Combination of Frequency Bands}
\label{sec_ilc}
We combine data from multiple frequency bands listed in Table~\ref{tab_survey_specs} using a scale-dependent linear combination~(LC) technique \citep{cardoso08, planck14_smica}. 
In harmonic space denoted by subscript $\ell$, this corresponds to
\begin{equation}
	S_{\ell m} = \sum_{i=1}^{N_\mathrm{bands}} w_\ell^i M_{\ell m}^i\, ,
\end{equation}
where $S$ is the desired sky signal (in our case the CMB), $\mathbf{w}_\ell$ is the weight, $\mathbf{M}_{\ell m}$ is the spherical harmonic transform of the input map, and the subscript $i$ runs over all the frequency bands. The multipole-dependent weights $\mathbf{w}_\ell$ are tuned in order to minimize the overall variance from noise and foregrounds. They are derived as 
\begin{equation}
	\mathbf{w}_\ell = \frac{\clinv \mathbf{A_s}}{\mathbf{A_s}^\dagger \clinv \mathbf{A_s}}\, .
\label{eq_ilc_weights_mv}
\end{equation}
\\
Here $\clcov$ is a $N_\mathrm{bands} \times N_\mathrm{bands}$ matrix containing the covariance of noise and foregrounds across different frequencies at a given multipole $\ell$; $\mathbf{A_s} = [1\ 1\ ..\ 1]$ is the frequency response of the CMB in different bands and is a $N_\mathrm{bands} \times 1$ vector. 
\refresponse{
%We model the foreground contribution using results from \citet{reichardt21} and the noise modeling is described in \S~\ref{subsec_survey_specs}.
The noise modeling is described in \S~\ref{subsec_survey_specs}.}

\refresponse{
The foreground signal receives contributions from both the galactic and extragalactic signals. 
In our baseline approach, we ignore the galactic foreground signals by introducing sky cuts but we perform tests to assess their impact using \texttt{pySM} simulations in \S\ref{sec_realworldeffects}.
The extragalactic foreground can be decomposed into signals from radio galaxies, dusty star forming galaxies (DSFG), tSZ and kSZ signals. 
We model these using measurements from \citet{reichardt21}. 
To this end, we assume a conservative point source masking threshold, similar to \citet{reichardt21}, and remove sources with flux above $S_{150}$ = 6 mJy. 
For the radio galaxies, we assume a spectral index of $\alpha_{\rm radio} = -0.76$ \citep{reichardt21} to model their frequency dependence. 
On the other hand, the DSFG contribution is split into clustering and Poisson components, which are both modeled using a modified blackbody model with $T_{\rm CIB} = 25 {\rm K}$, and $\beta_{\rm CIB-Clus} = 2.2$ and $\beta_{\rm CIB-Poi} = 1.5$ for the clustering and Poisson terms respectively. 
For the polarization power spectrum $EE$, we scale the above $TT$ power spectrum of sources as $C_{\ell}^{EE} = P^{2} C_{\ell}^{TT}$ assuming 
polarization fraction of $P=3\%$ ($P = 2\%$) for radio galaxies (DSFG) based on \citet{datta19, gupta19}. 
%Even though, this simple scaling assumes a non-zero $C_{\ell}^{TE}$ from sources, this has no impact on our results. 
We also remove the contribution from clusters detected at $\snr \ge 5$ which roughly corresponds to ${\rm M}_{500c} \gtrsim 10^{14} {\rm M}_{\odot}$ \citep[see Fig. 5 of][]{raghunathan22b}. 
For kSZ, we assume a flat $D_{\ell} = 3 \mu {\rm K}^{2}$ for all the bands. 
The tSZ and kSZ signals are assumed to be unpolarized, and hence we ignore the contribution from clusters on $C_{\ell}^{EE}$ and $C_{\ell}^{TE}$. 
}

\refresponse{We also note here that the instrumental bandpass of the individual frequency bands are not delta functions and are roughly $25$, $30$, and $50$ GHz for 95, 150, and 220 GHz bands respectively \citep{sobrin22}. The frequency responses of the detectors are calculated using a Fourier transform spectrometer. 
With these bandpass measurements, we calculate the effective frequencies $\nu_{\rm eff}$ for each of the foreground signals described above to compute their respective power spectra. 
For simplicity, we assume the values of $\nu_{\rm eff}$ to be the same as the ones reported in \citet{reichardt21} although the bandpasses of the current \sptnew{} receiver are slightly different \citep[see Fig. 7 of][]{sobrin22}. 
We checked the impact of the difference in the bandpasses, and it has negligible impact on the foreground modeling and the constraints reported in this work.
}

In Fig.~\ref{fig_noise_ilc_residuals}, we show the LC residuals for all three SPT-3G surveys as the thick black solid curve.
The top and bottom panels in the figure are for CMB temperature~($TT$) and polarization~($EE$) power spectra, denoted as gray curves.
In the absence of foregrounds, the LC residuals simply correspond to the inverse variance combination of noise from different frequency bands, which is almost true for $EE$ (bottom panels).

In Fig.~\ref{fig_powspec_sn}, we present signal-to-noise ($\snr$) per multipole for $TT/EE/TE/\phi\phi$ spectra for \sptnew{} Main (green) and \allspt{} (orange) surveys along with \planck{} (black). 
The power spectrum errors were calculated using the analytical formula (see Eq.\ref{eq_analytic_covariance}) \citep{knox95, jungman96a, zaldarriaga97c}. 
\planck{} has a higher $\snr$ on large scales because of the larger sky coverage. 
However, the $\snr$s of \mainfield{} and \allspt{} surveys are much higher than \planck{} on small scales. 
In particular, the $\snr$s for the \allspt{} survey will surpass \planck{} beyond $\ell  =$ 1800, 800, and 450 for $TT$, $TE$, and $EE$ respectively. 
For $\phi\phi$, the $\snr$ are better than \planck{} on all scales. 
As a result, we can expect significant improvement on cosmological parameters that control structure formation and that are sensitive to the damping tail from \sptnew{} compared to \planck{} as we show later in this work. 

\subsection{Analytic method}
\label{subsec_fisher_formalism}
%We also use the standard Fisher formalism to validate our forecasts using MUSE described above. 
Given a set of power spectra of the lensed CMB and the lensing potential, $X, Y \in [TT, EE, TE, \phi\phi]$, the Fisher matrix is given by,
\begin{equation}
    \mathcal{F}_{ij} = \sum_{X,Y} \sum_{\ell} \dfrac{\partial C_\ell^{X}}{\partial \theta_i} \cdot \left(\Sigma_\ell^{-1}\right)_{XY} \cdot \dfrac{\partial C_\ell^{Y}}{\partial \theta_j},
    \label{eq_fisher_matrix}
\end{equation}
where $\theta$s correspond to the cosmological parameters being constrained. The partial derivatives of the spectra with respect to the parameters $\theta$, denoted as $\partial C_\ell / \partial \theta_j$, are obtained using the finite difference method. $\Sigma_{\ell}$ is the covariance of the power spectra, given by Eq.(\ref{eq_analytic_covariance})
\begin{widetext}
        \begin{align}
    \Sigma_{\ell} = \dfrac{2}{\left(2\ell+1\right)f_{\rm sky}}
        \begin{pmatrix}
            (\tilde C_\ell^{TT})^2 & (C_\ell^{TE})^2 & \tilde C_\ell^{TT} C_\ell^{TE} & (C_\ell^{T\phi})^2\\
            (C_\ell^{TE})^2 & (\tilde C_\ell^{EE})^2 & \tilde C_\ell^{EE} C_\ell^{TE} & (C_\ell^{E\phi})^2\\
            \tilde C_\ell^{TT} C_\ell^{TE} & \tilde C_\ell^{EE} C_\ell^{TE} & \frac{1}{2}\left[(C_\ell^{TE})^2 + \tilde C_\ell^{TT}\tilde C_\ell^{EE}\right] & C_\ell^{T\phi}C_\ell^{E\phi}\\
            (C_\ell^{T\phi})^2 & (C_\ell^{E\phi})^2 & C_\ell^{T\phi}C_\ell^{E\phi} & (\tilde C_\ell^{\phi\phi})^2
        \end{pmatrix},
        \label{eq_analytic_covariance}
    \end{align}
\end{widetext}
where $f_{\textrm{sky}}$ is the fraction of the sky for the survey under consideration and 
\begin{equation}
    \tilde C_\ell^{X} = C_\ell^{X} + N_\ell^{X},
    \label{eq_cl_plus_nl}
\end{equation}
$X = \{TT, EE, \phi\phi\}$. $N_{\ell}^{TT}$ and $N_{\ell}^{EE}$ are the residual noise and foreground power after combining data from all the bands using the LC method. $N_{\ell}^{\phi\phi}$ is the lensing reconstruction noise derived using Quadratic Lensing Estimator \citep{hu02a}. In Table~\ref{tab_cosmo_parameters} we list the fiducial values of the cosmological parameters at which the Fisher matrix is evaluated and associated priors. 

\newcommand{\logAs}{{\rm ln}(10^{10}A_{s})}
\begin{deluxetable}{| c | c | c |}
\tabletypesize{\small}
\def\arraystretch{1.1}
\tablecaption{Fiducial values of the parameters and priors used in this work. All the applied priors are Gaussian with the widths given below in the table.}
\label{tab_cosmo_parameters}
\tablehead{
Parameter & Fiducial & Prior
}
\startdata
\hline
\multicolumn{3}{l}{\it $\lcdm$ parameters:}\\\hline
Amplitude of scalar & \multirow{2}{*}{3.044} & \multirow{7}{*}{-}\\
fluctuations $\logAs$ & & \\\cline{1-2}
Dark matter density $\omchsq$ & 0.1200 & \\\cline{1-2}
Baryon density $\ombhsq$ & 0.02237 & \\\cline{1-2}
Scalar spectral index $n_{s}$ & 0.9649 & \\\cline{1-2}
%Angular size of sound horizon & \multirow{2}{*}{1.04092} & \\
%at recombination $100\theta_{\ast}$ & & \\\hline
Hubble parameter $h$ & 0.6732 & \\\hline
Reionization optical depth $\taure$ & 0.0544 & 0.007 \\\hline
\multicolumn{3}{l}{\it Extensions:}\\\hline
Sum of neutrino masses $\summnu$ & 0.06 eV & \multirow{6}{*}{-} \\\cline{1-2}
Number of relativistic species $\neff$ & 3.044 & \\\cline{1-2}
Running of the spectral index $\nrun$ & 0 & \\\cline{1-2}
Spatial curvature $\omk$ & 0 & \\\cline{1-2}
Equation of state of dark energy $\wde$ & -1 & \\\cline{1-2}
Helium abundance $\yp$\tablenotemark{$^{\dagger}$} & 0.245 & \\\cline{1-2}
\enddata
\tablenotetext{\dagger}{Set from BBN (Big Bang nucleosynthesis) consistency when $\yp$ is not included in the extension.}
\end{deluxetable}

We calculate the Fisher matrix for each of the \sptnew{} surveys given in Table~\ref{tab_survey_specs} and combine them to obtain the final SPT-3G Fisher matrices as
\begin{align}
    \mathcal{F}_{\rm Summer} & = \mathcal{F}_{\rm Summer-A} + \mathcal{F}_{\rm Summer-B} + \mathcal{F}_{\rm Summer-C}\nonumber\\
    \mathcal{F}_{\rm Ext-4k} & = \mathcal{F}_{\rm Main} + \mathcal{F}_{\rm Summer}\nonumber\\
    \mathcal{F}_{\rm Ext-10k} & = \mathcal{F}_{\rm Main} + \mathcal{F}_{\rm Summer} + \mathcal{F}_{\rm Wide}\, .
    \label{eq_fisher_matrix_decompositions}
\end{align}
The parameter covariance matrices are then obtained by inverting the Fisher matrix.
We fix the $\ell$ ranges for all the SPT-3G surveys to the following for the Fisher matrix calculation: \mbox{$300 \leq \ell \leq 3500$} for $TT/TE$; \mbox{$300 \leq \ell \leq 4000$} for $EE$; and \mbox{$30 \leq L \leq 3500$} for $\phi\phi$. 
For lensing reconstruction, we use \mbox{$300 \leq \ell \leq 3500$} for T and \mbox{$300 \leq \ell \leq 4000$} for P. 
We ignore $\ell > 3500$ in $TT/TE$ because of the difficulties in modeling the extragalactic foreground signals in the temperature data. 
We have checked the correlations between the \mainfield{} and the contiguous \summerfield{} surveys and we have found that they are negligible. This is due to the fact that we do not include the large-scale modes $\ell \leq 300$.
Since the extragalactic foregrounds are largely unpolarized \citep{datta19, gupta19}, we can set a higher $\ell_{\rm max}$ for $EE$. 
We find that extending $\ell_{\rm max}^{EE} > 4000$ results in marginal improvements, indicating those modes are noise dominated, as can be inferred from Fig.~\ref{fig_powspec_sn}. 
%%%%%%%%%%%%%%%%%%%%%%%%%%%%%%%%%%%%%%%%%%%%%%%%%
\commenter{Since the extragalactic foregrounds are largely unpolarized \citep{datta19, gupta19}, we can set a higher $\ell_{\rm max}$ for $EE/TE$. 
While we also limit $\ell_{\rm max}^{TE} = 3500$, extending it further only improves our constraining power marginally. 
Similarly, extending $\ell_{\rm max}^{EE} > 4000$ also results in incremental improvements indicating that modes above $\ell \gtrsim 3500$ are noise dominated for $EE/TE$.}
%%%%%%%%%%%%%%%%%%%%%%%%%%%%%%%%%%%%%%%%%%%%%%%%%%%

\subsection{MUSE}
\label{sec_muse}

%\srini{Add few sentences mentioning we do not expect significant differences in parameter errors from MUSE; We only use MUSE to show that the comparison is useful for the data paper that uses MUSE and it has helped us understand few details.}

MUSE \citep{millea21_muse} is a general algorithm for approximate marginalization over arbitrary latent parameters which yields Gaussianized, asymptotically unbiased and near-optimal constraints on parameters of interest. 
In our application of MUSE, the parameters of interest are the unlensed $TT$, $TE$, $EE$, and $\phi\phi$ bandpowers, which are weighted averages of the respective power spectra across defined multipole bins. These bandpowers depend on the latent parameters, which are the pixel values of the lensing map and the unlensed CMB field maps.

As previously demonstrated by \citet{millea21} using \sptpol{} data, the standard Quadratic Lensing Estimator is sub-optimal for lensing reconstruction at noise levels below \mbox{$\Delta_{T} \sim 5 \ukarcmin$} \citep{hirata03a, hirata03b} and hence, using MUSE has an advantage in terms of the lensing reconstruction.
Another advantage of MUSE is that it performs optimal delensing on the lensed CMB maps and produces estimates of the unlensed power spectra, lensing potential power spectra and their joint covariance. This covariance naturally includes lensing-induced correlations between different bandpowers ($\ell-\ell^{\prime}$); and also between delensed CMB $TT/EE/TE$ and lensing $\phi\phi$ spectra \citep{peloton17, trendafilova23}. 

Despite the above advantages, we do not expect a significant improvement in the constraints from using MUSE over QE methods. 
This is because the lensing bandpower errors are dominated by sample variance over the angular scales for which MUSE reduces $\phi$ map noise, and the lensing-induced correlations are not important for the current noise levels. 
Moreover, the optimality of MUSE is non-negligibly better only for the \mainfield{} survey, and not for the others given their higher noise levels (see Table~\ref{tab_survey_specs}). 

\begin{figure*}
    \centering
\ifdefined\apjformat
    \includegraphics[width=0.8\textwidth, keepaspectratio]{lensing_noise_comb.pdf}
\else
    \includegraphics[width=0.8\textwidth, keepaspectratio]{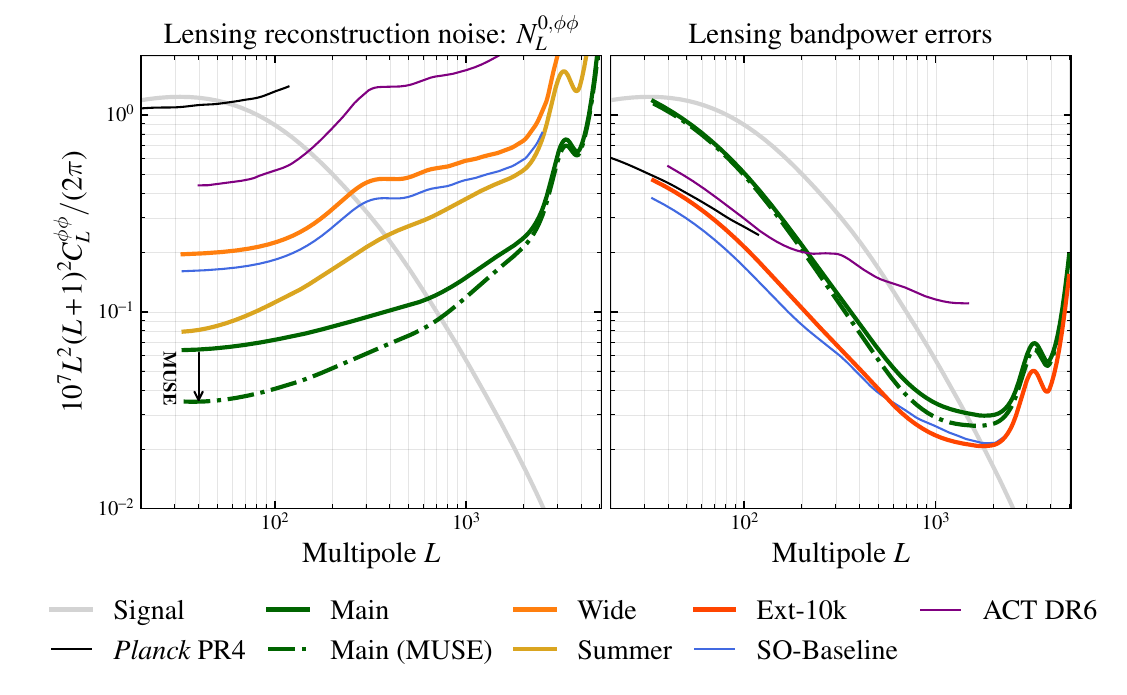}
\fi
    \caption{\textit{Left panel:} Lensing reconstruction noise $N_{L}^{(0, \phi\phi)}$ power spectra for the \sptnew{} surveys along with \planck{}, SO and ACT. 
    For the \mainfield{} survey, we show two curves: solid green for QE and dash-dotted green for MUSE.
    For \mainfield{} survey noise levels, lensing noise from MUSE is roughly $\times2$ better than QE at large scales. 
    %For the polarization-only lensing reconstruction (not shown), MUSE is $\times2.5$ better than QE.
    For other surveys, the difference between MUSE and QE is negligible and hence, not shown. 
    \textit{Right panel:} Bandpower errors for the lensing power spectrum (gray) from different surveys. 
    While the lensing reconstruction noise $N_{L}^{(0, \phi\phi)}$ is lower on large scales for MUSE, the bandpower errors are not significantly different as they are dominated by the sample variance.
    }
\label{fig_lensing_comb}
\end{figure*}

%\srini{Address KW's comment about moving N0 improvement.}
%The main motivation behind using MUSE for forecasting is to understand the differences, if any, between the analytic covariance in Eq.(\ref{eq_analytic_covariance}) and MUSE, as MUSE is currently being used as one of the analysis pipelines for the \mainfield{} survey data from the first two observation years. 
The optimality of the MUSE estimates is useful for other purposes like delensing the lensing-induced B-modes and cross correlations, at least for the \mainfield{} survey with its low noise levels (see Table~\ref{tab_survey_specs}). 
We demonstrate this in Fig.~\ref{fig_lensing_comb} which shows the lensing reconstruction noise $N_{L}^{(0, \phi\phi)}$ for different surveys. 
For the \mainfield{} survey shown in green, the lensing reconstruction noise goes down by roughly $\times2.5\ (\times 1.9)$ for polarization-only (including temperature information) lensing estimators when we replace QE (solid) with MUSE (dash-dotted). 
As mentioned above, the difference between QE and MUSE is negligible for other surveys and hence, not shown. 
%\srini{The discussion about N0 improvement and Fig.4 might be moved to the appendix during CWR.}

In MUSE, we express the lensing and unlensed CMB power spectra as scaled versions of the fiducial power spectra, which, for our purposes, are based on the \refresponse{\planck{} 2018 spectra \citep{planck18-6}}. This is given by
\begin{equation}
C_\ell = A_\ell C_\ell^{\text{fid}}\, ,
\end{equation}
where $C_\ell^{\text{fid}}$ represents the fiducial power spectrum, and $A_{\ell}$ is the scaling factor applied to this fiducial spectrum. Here, $C_\ell$ is used as a general notation that encompasses all types of power spectra, including unlensed $TT/ TE/ EE$ and $\phi\phi$. 

In our application of MUSE, we do not estimate the power at each individual multipole but instead bandpower parameters, $A_b$, that each scale a fiducial power spectrum across a band of width $\Delta \ell = 100$. 
As a result, MUSE produces a covariance matrix of these scaling coefficients $\Sigma_{A_b} \equiv \left< A_{b} A_{b}^{*} \right>$. The $A_{b}$ can be understood as the binned version of $A_\ell$. 

As described below, this covariance is subsequently utilized to construct the cosmological parameter-Fisher matrix, from which we derive the constraints on parameters. 
We provide more details about MUSE in the appendices. 
Specifically, we describe the derivation of the MUSE covariance matrix in Appendix~\ref{appendix_muse_cov}, the simulations used for estimating the covariance matrix in Appendix~\ref{appendix_muse_sims}, and the lensing reconstruction noise in Appendix~\ref{appendix_muse_lensing_noise}.

The standard Fisher formula described in the previous section Eq.(\ref{eq_fisher_matrix}) can be adapted to incorporate the MUSE covariance matrix $\Sigma_{A_b}$. 
Let the covariance matrix of the power spectra be represented by $\Sigma_{C_\ell}$ and the covariance matrix of the amplitudes be represented by $\Sigma_{A_\ell}$, then the Fisher matrix can be written as,

\begin{align}
    \mathcal{F}_{ij} &= \sum_{XY} \sum_{\ell \ell^{\prime} } \dfrac{\partial C_\ell^{X}}{\partial \theta_i}  \left(\Sigma_{C_\ell}^{-1}\right)_{XY}^{\ell \ell^{\prime}} \dfrac{\partial C_{\ell^{\prime}}^{Y}}{\partial \theta_j} \\
     &= \frac{1}{C^{\text{fid}}_\ell}\frac{\partial C_\ell}{\partial \theta_i} \cdot \left(\Sigma_{A_\ell}^{-1}\right) \cdot \frac{1}{C^{\text{fid}}_\ell}\frac{\partial C_\ell}{\partial \theta_{j}} \\
     &\approx \dfrac{\partial A_b}{\partial \theta_i} \cdot \left(\Sigma_{A_b}^{-1}\right) \cdot \dfrac{\partial A_b}{\partial \theta_j}.
\end{align}

For ease of notation, we drop the summation symbols and transition to using matrix products in the second line. In the last line we assume that the Fisher matrix is approximately equivalent to its binned version. We obtain the derivatives of $A_b$ with respect to the cosmological parameters via a minimum-variance binning of  $\partial C_\ell/\partial \theta$ given by
\begin{equation}
\dfrac{\partial A_b}{\partial \theta} = \dfrac{1}{\sum_{\ell \in b}\dfrac{1}{\sigma_{A_\ell} ^2}} \sum_{\ell \in b} \dfrac{1}{\sigma_{A_\ell} ^2} \dfrac{\partial C_\ell/\partial \theta}{C_\ell^{\text{fid}}}\, ,
\end{equation}
where we use an analytic ansatz for the weighting function $\sigma_{A_\ell}^2$. 
For $X \in [TT, EE, \phi\phi]$ this is given by% Eq.(\ref{eq_sigma_Al}) 
\begin{equation}
\sigma_{A_\ell^{X}}^2 
= \frac{\sigma^2_{C_\ell^{X}}}{(C_\ell^{X, \text{fid}})^{2}}
\equiv \frac{2}{2\ell +1} \frac{(C_\ell^{X} + N_\ell^{X})^2}{(C_\ell^{X, \text{fid}})^{2}}
\label{eq_sigma_Al}
\end{equation} and for $TE$ it is
\begin{equation}
\sigma_{A_\ell^{TE}}^2 
= \frac{1}{2\ell +1} \frac{(C_\ell^{TE})^2 + (C_\ell^{TT}+N_\ell^{TT})(C_\ell^{EE}+N_\ell^{EE})}{(C_\ell^{TE, \text{fid}})^{2}}.
\label{eq_sigma_Al_te}
\end{equation}

\begin{comment}
For $TT, EE, \phi\phi$ spectra, this is given by the equation below, with an analogous term for $TE$ (See equation \ref{Sigma_l}). 
\begin{equation}
\sigma_{A_\ell}^2 
= \frac{\sigma^2_{C_\ell}}{(C_\ell^{\text{fid}})^2}
\equiv \frac{2}{2\ell +1} \frac{(C_\ell + N_\ell)^2}{(C_\ell^{\text{fid}})^2},
\end{equation}

\begin{align}
    \sigma_{A_\ell}^2 &= \frac{\sigma^2_{C_\ell}}{(C_\ell^{\text{fid}})^2} \\
    &\equiv
    \begin{cases}
        \frac{2}{2\ell +1} \frac{(C_\ell + N_\ell)^2}{(C_\ell^{\text{fid}})^2},\\
        & \text{for } TT, EE, \phi\phi \\
        \frac{1}{2\ell +1} \frac{(C_\ell^{TE})^2 + (C_\ell^{TT}+N_\ell^{TT})(C_\ell^{EE}+N_\ell^{EE}) }{(C_\ell^{\text{fid}})^2},\\ 
        & \text{for } TE
    \end{cases}
\end{align}
\end{comment}
%The traditional estimators usually assume full-sky or periodic flat-sky boundary conditions without masking. But the presence of masking can induce mode couplings that can break some of these simple approximations. 

\begin{figure}
    \centering
\ifdefined\apjformat
    \includegraphics[width=0.48\textwidth, keepaspectratio]{museVsStandard.pdf}
\else
    \includegraphics[width=0.48\textwidth, keepaspectratio]{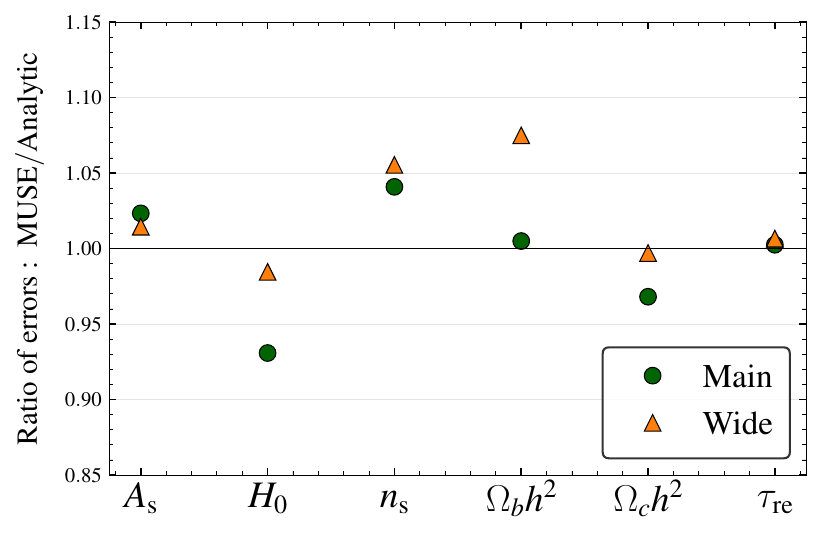}
\fi
    \caption{Ratio of the \llcdm\ parameter constraints derived from MUSE to those derived from analytical covariance matrix for the \mainfield{} survey (dark green circle) and the \extfield{} survey (orange triangle). We note that they agree with each other within 10\%.}
    \label{fig_museVsStandard}
\end{figure}

%\srini{After binning, other parameters might match MUSE but H0 can become more discrepant ($>10\%$). So KP to run analytic with unlensed spectra to see if that improves the analytic to match with MUSE; Also may be pick MUSE's N0 and rerun the forecasts.}
In Fig.~\ref{fig_museVsStandard}, we show the ratio of parameter constraints obtained from MUSE and the analytic covariance for \sptnew{} Main (green) and the \sptnew{} Wide (orange) surveys. We find that these methods are consistent with each other within a 10\% margin, which we deem sufficient to satisfy our validation criteria. 

There are a number of reasons to expect these errors to differ to some degree, which we already brought up when describing the advantages of MUSE. 
Another reason is that the analytic results are not binned, but we found that for both the Wide and Main surveys, binning to $\Delta \ell = 100$ only leads to at most 2\% degradation in constraints. 
The largest difference is the 8\% tighter constraint on $\Omega_b h^2$ from the analytic forecast for the \extfield{} survey. 
This is because the analytic estimate does not include the $\ell$-by-$\ell$ lensing-induced correlations among different CMB multipoles, and that can lead to overly-optimistic constraints.
By artificially removing off-diagonal elements from the MUSE covariance matrix\footnote{Specifically we did this by inverting it, removing the off-diagonals, and inverting back}, we were able to largely reproduce this reduced analytic error for both $\Omega_b h^2$ and $n_s$. 

%The minor discrepancies observed  between the two methods can be attributed to the following factors: 1. MUSE employs lensing noise curves derived from optimal delensing instead of the quadratic estimator used in the analytic method. 2. MUSE uses an estimate of the delensed spectra in contrast to the lensed CMB spectra used in the analytic method. 3. The covariance from MUSE takes into account the correlations across the different angular scales as well as those between lensing and CMB power spectra. Additionally, a fourth distinction arises from the fact that MUSE uses binned spectra while the analytic method uses unbinned ones. However, we verified that the impact of binning is minimal, with the most significant effect observed in $H_0$ constraint, which degrades by 2\%. 

\subsection{Combining with \planck{} data}
\label{subsec_spt_planck}
Comparison of the SPT-only \llcdm\ constraints with those from \planck{}, as we will see, will enable a powerful test of the \llcdm\ model.
If these tests are consistent with \llcdm\, then we will be interested in using the combined SPT-3G+\planck{} data to estimate \llcdm\ parameters, as \planck{} provides complementary information on large angular scales. 
Consistent or not, we will also be interested in using the combined data to constrain \llcdm\ extensions. Hence, we report the expected constraints from the combined datasets under the assumption of \llcdm\ as well as under the assumption of various extensions. %For extensions to \llcdm\, we only quote the combined SPT and \planck{} forecasts and do not report SPT-only predictions.

%We primarily focus on SPT-only constraints for the standard \llcdm\ parameters. This is to probe any systematic shifts in the parameters between SPT and other experiments like \planck{}. However, combining \planck{} with SPT is also interesting as \planck{} improves our sensitivity to large scales by virtue of its extensive sky coverage and the absence of atmospheric noise. Subsequently, we also report the expected constraints on \llcdm\ parameters by combining SPT with \planck. For extensions to \llcdm\, we only quote the combined SPT and \planck{} forecasts and do not report SPT-only predictions. 

The standard approach to obtain this combination is to add the Fisher matrices from SPT-3G and \planck{} similar to Eq.(\ref{eq_fisher_matrix_decompositions}) as
\begin{align}
    \mathcal{F}_{{\rm SPT} + \planck} & = \mathcal{F}_{\rm \widehat{SPT}} + \mathcal{F}_{\widehat{\planck}}
    \label{eq_fisher_matrix_spt_planck},
\end{align}
where $\mathcal{F}_{\rm \widehat{SPT}}$ is the modified SPT-3G Fisher matrix obtained by taking the inverse variance combination of SPT-3G and \planck{} noise in the field under consideration; and $\mathcal{F}_{\widehat{\planck}}$ is the modified \planck{} Fisher matrix after scaling to remove the overlapping sky region between SPT-3G and \planck. 
For example, if $\fsky = 1$ for $\mathcal{F}_{\planck}$, then for $\mathcal{F}_{\widehat{\planck}}$, we use $\fsky = 1 - \fsky^{\rm SPT}$. 

However, obtaining Eq.(\ref{eq_fisher_matrix_spt_planck}) is non-trivial as it requires us to know the sky fraction used by \planck, which varies for different bands and power spectra \citep{planck18-6}. 
Instead, we follow a different approach which involves two steps: First, we derive the \planck{} Fisher matrix directly from the publicly available \planck{} PR3 chains \citep{planck18-6}; second, we combine the above \planck{} Fisher matrix with SPT-3G Fisher matrix, obtained using the analytic method. We modify the SPT-3G Fisher matrix by removing the large-scale modes from it,
%, rather than scaling the \planck{} Fisher matrix, 
to avoid double counting information in the overlapping region.

\begin{figure}
\ifdefined\apjformat
    \includegraphics[width=0.45\textwidth, keepaspectratio]{Bandpower_error_ratios_spt_planck.pdf}
\else
    \includegraphics[width=0.45\textwidth, keepaspectratio]{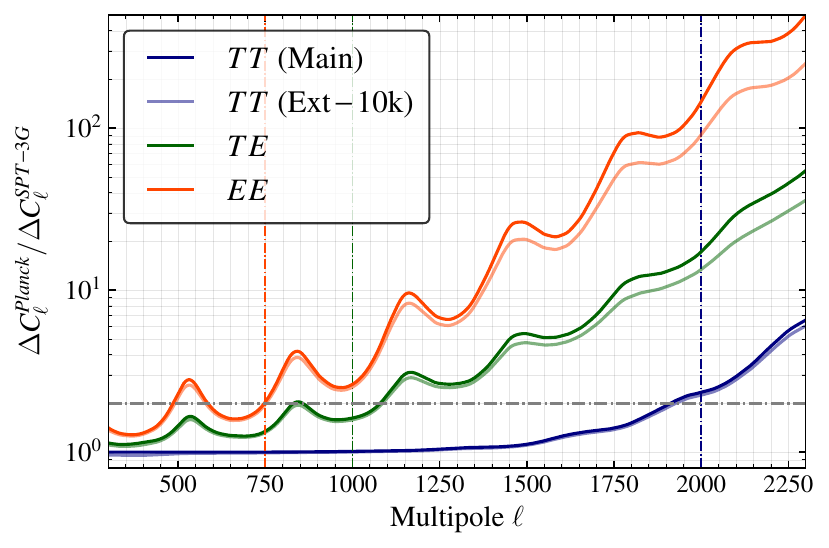}
\fi
\caption{Ratio of power spectrum errors for \planck{} to those of \sptnew{} for the same sky fraction.
$\ell_{min}$, the multipole value at which we switch from \planck\ to \sptnew\ in combined constraints (shown as the vertical dash-dotted lines), is chosen to be where this ratio roughly crosses $2$.
The blue, green, and orange colors correspond to $TT, TE, EE$ respectively. 
For $\phi\phi$ this ratio is greater than $2$ for all the multipoles used in the forecast, and therefore it is not shown.
The darker curves correspond to the \mainfield{} survey whereas the lighter curves represent the \allspt{} survey. 
The choice of $\ell_{\rm min}$ does not change significantly for different SPT-3G surveys.}
\label{fig_bandpower_ratio_spt_planck}
\end{figure}
 We use \texttt{BASE\_PLIKHM\_TTTEEE\_lowL\_lowE\_lensing}\footnote{\planck{} chains are downloaded from \url{https://wiki.cosmos.esa.int/planck-legacy-archive/index.php/Cosmological_Parameters}} chains for the first step unless specified otherwise. These chains, which contain samples from the probability distribution of cosmological parameters, are used to construct a parameter covariance matrix. By inverting this covariance matrix, we obtain the \planck{} Fisher matrix.
For the second step, we restrict the information we use from SPT power spectra to multipoles above some $\ell_{\rm min}$ or $L_{\rm min}$ threshold, set by the approximate multipole at which the $\snr$ for \planck{} is $\times2$ lower than that for SPT in the same patch. 
We find that the choice of $\ell_{min}$ does not change significantly for different SPT-3G surveys.
As can be seen from Fig.~\ref{fig_bandpower_ratio_spt_planck}, which shows the ratio of \planck\ $TT$, $TE$, and $EE$ power spectrum uncertainties to those projected for \sptnew{}, this threshold is roughly 2000, 1000, and 750 for $TT$, $TE$, and $EE$ respectively, and we adopt these values for the joint forecasting. 
For lensing power spectra, we set $L_{\rm min} = 30$ (the same as our fiducial setting) because \planck's lensing $\snr$ is much lower than SPT on all scales, as can be seen from Fig.~\ref{fig_powspec_sn}. 
%%%%%%%%%%%%%%%%%%%%%%%%%%%%%%%%%%%%%
%Replaced by Tom's suggestion
\commenter{
For the second step, we set $\ell_{\rm min} = 2000, 1000, 750$ for $TT, TE, EE$ and $L_{\rm min} = 30$ for $\phi\phi$ power spectra respectively when constructing the SPT Fisher matrices. 
We set the above $\ell_{\rm min}$ thresholds to be the multipole at which the $\snr$ for SPT is about $\times2$ higher than that of \planck{} in the same patch. 
In Fig.~\ref{fig_bandpower_ratio_spt_planck}, we present the ratio of signal to noise ratio ($\snr$) of \planck{} and SPT for different power spectra ($TT$ in blue, $EE$ in orange, and $TE$ in green) within the SPT patch. 
Here, $\snr$ is the ratio of individual power spectra over their errors. 
In the above figure, the opaque curve corresponds to the \mainfield{} survey while the semi-transparent curve is for the \allspt{}. 
As we can see, the thresholds does not change significantly for different SPT-3G patches. 
For lensing power spectra, we set $L_{\rm min} = 30$, same as our fiducial setting, because \planck's lensing $\snr$ is much lower than SPT on all scales as can be seen from Fig.~\ref{fig_powspec_sn}.
}
%%%%%%%%%%%%%%%%%%%%%%%%%%%%%%%%%%%%%
However, we find our forecasts to be insensitive to the choice of $L_{\rm min}$. 
For example, with $L_{\rm min} = 400$ for SPT, our constraints only weaken by $\lesssim 5\%$. 

We validate this approach of using $\ell_{\rm min}$ cuts 
rather than the $\fsky$ scaling in Eq.(\ref{eq_fisher_matrix_spt_planck}) 
by creating a toy model for a \planck-like Fisher matrix with \mbox{$\fsky=0.5$}. 
For this toy model, we use the following multipole ranges: \mbox{$\ell \in [2,2000],\ [30,2000],\ [30,2000]$} respectively for $TT$, $TE$, and $EE$ and $L \in [30,400]$ for $\phi\phi$. 
Next, we compare the Fisher forecasts from Eq.(\ref{eq_fisher_matrix_spt_planck}) and the $\ell_{\rm min}$-cut approach discussed above. 
We find the results from the $\ell_{\rm min}$ cut approach to be $\lesssim 8\%$ worse than the traditional method. 
Given that the two methods agree well and given that our $\ell_{\rm min}$-cut-based forecasts are slightly on the conservative side, we choose to use this approach for the rest of this work.

%--------------------------------------------------------------------------%
\section{Results}
\label{sec_results}

\begin{figure}
    \centering
    \ifdefined\apjformat
        \includegraphics[width=\linewidth, keepaspectratio]{fisher_fom.pdf}
    \else
        \includegraphics[width=\linewidth, keepaspectratio]{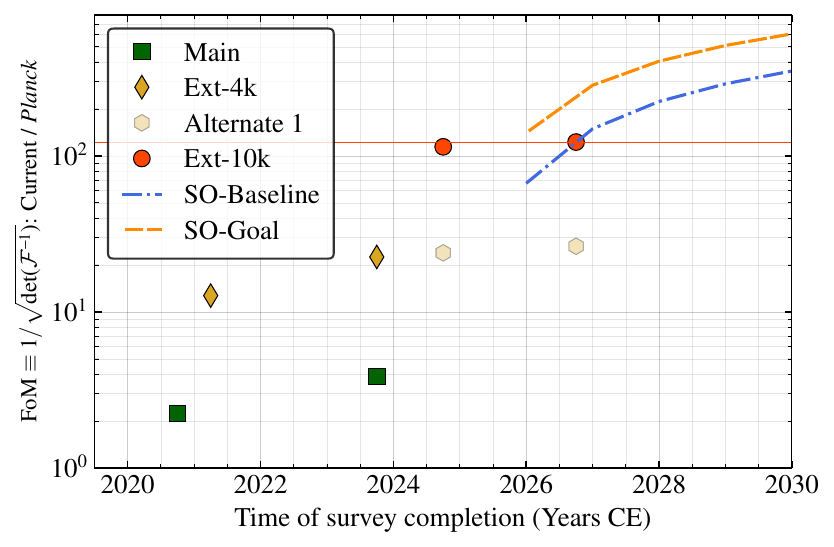}
    \fi
    \caption{Figure of merit (FoM) for \llcdm\ compared to \planck: 
    We present the FoM for \mainfield{} (green), Ext-4k (yellow), and \allspt{} (red). 
    The transparent yellow circle represents the improvement in \mainplussummerfield{}, if we were to continue the \mainfield{} survey during 2024 as opposed to conducting the \extfield{} survey. 
    The evident increase in FoM with the new strategy emphasizes its potential efficacy over continuing the \mainfield{} survey for cosmological parameter constraints. For reference, we also show the FoM expected for the two SO configurations (Baseline in blue and Goal in orange) under the assumption that the observations will begin in early 2025.
    }
    \label{fig_fom}
\end{figure}

\iffalse{
\begin{figure*}
    \centering
    %\includegraphics[width=0.9\textwidth, keepaspectratio]{figures/lcdm_constraints_spt_vs_planck.pdf}
    \ifdefined\apjformat
        \includegraphics[width=0.9\textwidth, keepaspectratio]{lcdm_constraints_spt_and_sptplusplanck_vs_planck.pdf}
    \else
        \includegraphics[width=0.9\textwidth, keepaspectratio]{figures/lcdm_constraints_spt_and_sptplusplanck_vs_planck.pdf}
    \fi
    %\includegraphics[width=0.9\textwidth, keepaspectratio]{figures/lcdm_constraints_spt_vs_planck_single_panel.pdf}
    \caption{Constraints on the standard \llcdm\ parameters expected from the three SPT surveys (\mainfield{} in green, \mainplussummerfield{} in yellow, and \allspt{} in red) compared to \planck{} (grey shaded region). We expect significant improvements on all the parameters except $n_{s}$ and $\taure$ which depend on the largest scales that ground-based surveys cannot easily measure.}
    \label{fig_lcdm_constraints_spt_vs_plank}
\end{figure*}
}\fi

\begin{figure}
    %\centering
    \ifdefined\apjformat
        \includegraphics[width=0.47\textwidth, keepaspectratio]{lcdm_constraints_spt_vs_planck_single_panel.pdf}
    \else
        \includegraphics[width=0.47\textwidth, keepaspectratio]{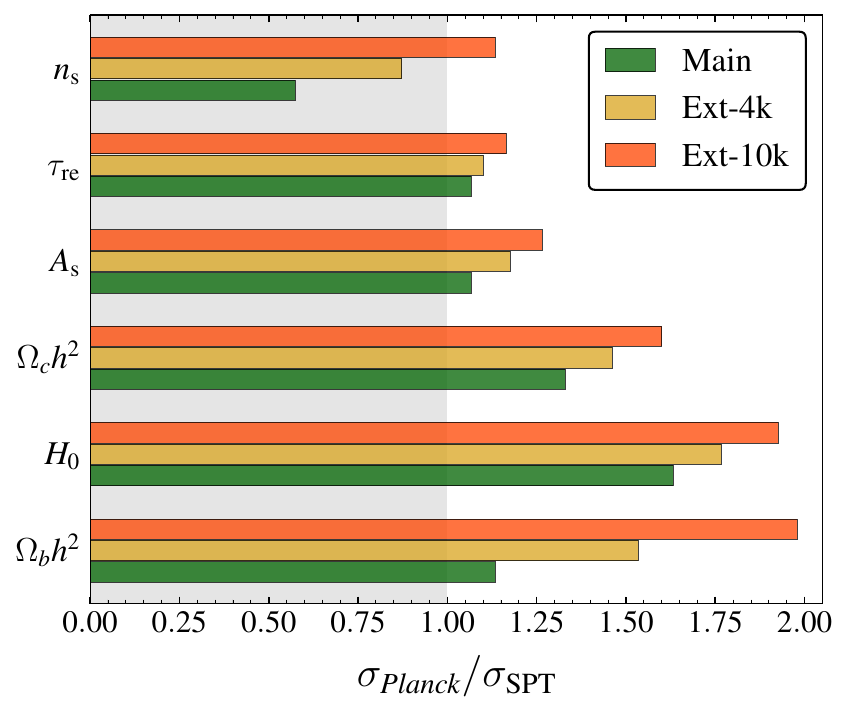}
    \fi
    \caption{Constraints on the standard \llcdm\ parameters expected from combinations of the three SPT-3G surveys (\mainfield{} in green, \mainplussummerfield{} in yellow, and \allspt{} in red) compared to \planck{} (grey shaded region). We expect significant improvements on all the parameters except $n_{s}$ and $\taure$ which depend on the largest scales that ground-based surveys cannot easily measure.
}
    \label{fig_lcdm_constraints_spt_vs_plank}
\end{figure}

In this section, we begin by presenting our forecasts on cosmological parameters using all three \sptnew{} surveys, assuming the 6-parameter \llcdm\ model. Subsequently, we project the extent to which combined \sptnew{} and \planck{} data can constrain both single- and double-parameter extensions to the \llcdm\ model. We also qualitatively investigate how well \sptnew{} data can differentiate between \llcdm\ and alternative models that have been proposed in the literature to address current cosmological tensions \citep[for example,][]{cyrracine21, schoneberg22, meiers23, hughes23, khalife23, schoneberg2022}.
%We emphasize that 
The models we consider are fully consistent with current CMB and lensing power spectra measurements but make different predictions, from those of \llcdm, for signals that can be measured well by \sptnew{}.

%In this section we describe \sptnew{}'s capability for testing the \llcdm\ cosmological model and exploring extensions in three distinct ways. First, we demonstrate that \sptnew{} is capable of imposing constraints on the \llcdm\ parameters that are tighter than those achieved from the \planck{} data. These constraints are obtained from largely independent signals than those from \planck{}, thus providing a robust consistency test for the \llcdm\ model. Second, we exhibit how incorporating \sptnew{} data with existing \planck{} data enhances the constraints on the \llcdm\ model and its simple one- and two-parameter extensions. Finally, we leverage \sptnew{} data to differentiate between the \llcdm\ model and alternative models proposed to address current cosmological tensions. Such models are fully consistent with the current CMB power spectrum and lensing data and \sptnew{} is capable of differentiating these models.

%%%look_into_this_later \srini{To delve into the detailed physics of these parameters, we encourage readers to consult the CMB-S4 science book \citep{abazajian16}. Beyond parameter-level cross-examinations and discovery potential, the comprehensive dataset from the \sptnew{} surveys paves the way for robust systematic error testing.  Given the immense implications of any departures from \llcdm\, it is crucial to validate experimental results, particularly as we advance to higher-precision polarization measurements and extend temperature measurements to smaller angular scales.}
%--------------------------------------------------------------------------%

\subsection{Forecasts of constraints on \llcdm\ parameters} %\texorpdfstring{\llcdm\}{LCDM} parameters}
\label{sec_lcdm_constraints}

%In this section we forecast \sptnew{}'s capability to constrain the \llcdm\ parameters, namely baryon density $\ombh$, cold dark matter density $\omch$, Hubble constant $H_0$, optical depth $\taure$, amplitude and spectral index of primordial fluctuations $A_s$ and $n_s$. We also include a prior on $\sigma(\taure)=0.007$ from \planck{} \citep{planck18-6}. We use the MUSE method to obtain these constraints. We also compare these constraints against the traditional Fisher matrix approach and find them to be in good agreement as shown in figure~\ref{fig_museVsStandard}. We show the projected Figure of Merit (FoM) value in figure~\ref{fig_fom} which is a inversely proportional to the 6-dimensional parameter space volume with respect to the current best constraints from \planck{}. We also include the projected FoM ratios for the SO survey. It is interesting to note that SO will reach \sptnew{}’s level roughly at the end of their second year. 

%The forecasted constraints on the \llcdm\ parameters from \mainfield{}, \extfield{}, and \allspt{} are shown in figure~\ref{fig_lcdm_constraints_spt_vs_plank} and also tabulated in table \ref{tab_lcdm_param_table}. In fig \ref{fig_lcdm_constraints_spt_vs_plank} we plot the ratio of 1$\sigma$ standard error on the parameters from the \planck{} results to those forecasted from \sptnew{}. The gray shaded region corresponds to the current best constraints from \planck{}. 

%\lcdmtablebothmuseandfisher{}

We start first with forecasts of constraints on the six parameters of the \llcdm\ model, namely baryon density $\ombh$, cold dark matter density $\omch$, Hubble constant $H_0$, optical depth to reionization $\taure$, amplitude (defined at the pivot scale \mbox{$k_0 = 0.05$ Mpc$^{-1}$}) and spectral index of scalar fluctuations, $A_s$ and $n_s$. 
For these constraints we fix the neutrino mass to the minimal value of $0.06$ eV.
We also include a prior on $\sigma(\taure)=0.007$ from \planck{} \citep{planck18-6}.
We use the MUSE approach as described in \S~\ref{sec_muse} to derive these constraints, and remind the reader that there is agreement between MUSE and the traditional analytic approach to better than 10\%, as shown in Fig.~\ref{fig_museVsStandard}.

As we will see below, \sptnew{} can produce constraints on \llcdm\ parameters that are tighter than those we have from \planck.
Since these constraints are obtained from largely different signals than those from \planck{}, they provide a robust consistency test for the \llcdm\ model.
As mentioned earlier in \S~\ref{subsec_spt_planck}, we also present \llcdm\ forecasts by combining \sptnew{} and \planck. 

In Fig.~\ref{fig_fom}, we present the projected Figure of Merit (FoM), which is inversely proportional to the volume of allowed 6-dimensional \llcdm\ parameter space. It is calculated as 
%\[{\mathbf{FoM}} = \sqrt{ 1/ \det \Sigma }\]
\begin{equation}
\label{eq_fom}
{\rm FoM} = \sqrt{ 1/ {\rm det}(\mathcal{F}^{-1}) }
\end{equation}
for different \sptnew{} surveys (\mainfield{} in green, \mainplussummerfield{} in yellow, and \allspt{} in orange-red) along with SO (blue). 
The FoM is inversely proportional to the 6-dimensional volume of the 68\% confidence region.\footnote{In our Gaussian approximation, this proportionality holds independent of whether it is 68\%, 95\% or any other such threshold for defining the confidence region} We plot it divided by the FoM for \planck. 

As is evident from the figure, \sptnew{} surveys significantly increase the FoM relative to that from \planck. 
From just the first two years (2019-2020) of \mainfield{} survey data, the FoM increases by $\times2.5$ over \planck, and that jumps to $\times13$ with the addition of the first two years of \summerfield{} data. Both of these are further improved by roughly $\times1.7$ with the inclusion of \mainfield{} and \summerfield{} survey data through 2023.
%%%%%%%%%%%%%%%%%%%%%%%%%%%%%%%%%%%%%%%%
%Replaced by Tom's suggestion
\commenter{We achieve $\times13$ ($\times2.5$) reduction just from the first two years of \mainplussummerfield{} (\mainfield) whose analysis is ongoing and close to completion. This is shown as a yellow (green) circle. 
These improve roughly by $\times1.7$ with the data from observations we have already completed (2019-2023) as shown by the second yellow (green) circle.
}
%%%%%%%%%%%%%%%%%%%%%%%%%%%%%%%%%%%%%%

The figure also demonstrates the benefit of expanding sky coverage rather than continuing to go deeper on the \mainfield{} survey patch. One can see this from the minimal improvement to \mainplussummerfield{} ($\times1.1$) that we anticipate if we continue with the \mainfield{} survey patch observations for the 2024 winter, indicated by the semi-transparent yellow circle. 
On the other hand, switching to the \extfield{} observations in the 2024 season improves the constraining power by $\times 5$ compared to \mainplussummerfield{} and by $\times130$ compared to \planck. 
%\sout{This improvement is due to the increased sky coverage and it is denoted as the red circle. 
%Note that the red circle denotes the constraining power from the full \allspt{} and not just from the \extfield{} survey.}
%We do not notice a significant improvement in the parameter volume when we observe the \extfield{} patch for one more year and hence during 2025-26 season, \sptnew{} will switch back to the \mainfield{} survey to further improve the noise levels and subsequently the lensing map to facilitate delensing for B-modes.\\

Were we to observe
the \extfield{} patch for one more year we would not notice a significant increase in the FoM. Instead of doing this we plan to switch back to the \mainfield{} survey during the 2025 and 2026 winter seasons. 
Further noise reduction in the \mainfield{} survey will not increase the FoM much, as can be seen in Fig.~\ref{fig_fom}, but will significantly improve the measurements of the secondary CMB anisotropies including lensing, which will further improve the subtraction of lensed B-modes and hence, the search for B-modes from primordial gravitational waves \citep{bicep2keck21}.

\lcdmtable{}
In Table~\ref{tab_lcdm_param_table} we list our forecasts of \llcdm\ parameter errors for different \sptnew{} datasets. 
We also present them in Fig.~\ref{fig_lcdm_constraints_spt_vs_plank}, where the bars show the ratios of the values of $\sigma(\theta)$ from \planck{} to those from SPT. 

Let us begin by noting two things relevant for interpretation of our
$A_s$ and $\taure$ results.
First, the calibration of 
SPT-3G maps relies on data from \planck{} \citep{dutcher21}, and, consequently, SPT-3G constraints on the normalization parameter $A_{s}$ are not entirely independent of \planck.
%the CMB maps generally relies on data from \planck{} \citep{dutcher21} which will affect the normalization parameter $A_{s}$. Consequently, the constraints on $A_{s}$ are not entirely independent of \planck.
Second, for SPT-only constraints in this work, we include a \planck-like prior on the optical depth, $\sigma(\taure) = 0.007$, since $\taure$ is primarily constrained by the large-scale $\ell \lesssim 10$ reionization bump in $EE/TE$, not accessible by SPT. All the $\taure$ constraints are dominated by this prior.
These two points are not unrelated since the combination \mbox{$A_{s}\ e^{-2\taure}$} is very tightly constrained by CMB data at $\ell > 20$, so improvements in $\taure$ determination translate into improvements in $A_s$ determination, and vice versa. The small improvements that we do see in $\taure$ are due to the dependence of the amplitude of gravitational lensing on $A_s$, an amplitude unaffected by $\taure$. 
%However, these scales are omitted in the SPT data due to the filtering strategies adopted to mitigate excess noise along the scan direction. 

%Bearing in mind the caveats for $A_s$ and $\taure$, 
Focusing on the other four parameters, we find that the SPT-only constraints are generally comparable to, or better than, those obtained from \planck{} alone, for each of our three SPT-3G survey combinations.
For the case of \allspt{}, the improvements in $H_0$ and $\Omega_bh^2$ are nearly a factor of 2, and for $\Omega_c h^2$ they are just more than a factor of 1.5.
%This improvement is especially significant when considering the comparison of $H_{0}$ constraints from different scales, as probed by \planck{} and SPT.
For $n_{s}$, the relative errors from \mainfield{} survey and \mainplussummerfield{} survey are larger compared to those from \planck{} due to \planck's significantly broader multipole range. Nonetheless, the \allspt{} survey still manages to achieve slightly tighter constraints on $n_s$ than what we have from \planck.\\

%SO results for reference.
%SO-Baseline: FOM = 1.1729877838806492e+28; so_baseline_constraints = {'As': 2.1658200150842984e-11, 'h': 0.26675474816345607, 'ns': 0.0029954115279581814, 'ombh2': 5.909247162970899e-05, 'omch2': 0.0006940985538992085, 'tau': 0.005843945275405404}

%so_baseline_constraints_with_highertauprior = {'As': 2.5301485812922746e-11, 'h': 0.3041040336195913, 'ns': 0.003161575661733512, 'ombh2': 5.9097660559355926e-05, 'omch2': 0.0007950492456299664, 'tau': 0.0068650111552127}

%SO-Goal: FOM = 2.0282131576901477e+28; so_goal_constraints = {'As': 2.1218921980934253e-11, 'h': 0.25412315708878497, 'ns': 0.0028485888092844103, 'ombh2': 5.1267416668502566e-05, 'omch2': 0.0006689668479048663, 'tau': 0.00576499837079744}
%SO: FOM = xxx; so_original_constraints = {'As': 3e-11, 'h': 0.3, 'ns': 0.002, 'ombh2': 0.00005, 'omch2': 0.0007, 'tau': 0.007}
%for ppp in so_original_constraints:
%    print(ppp, so_original_constraints[ppp] / so_baseline_constraints[ppp])

\subsubsection{Comparison with SO} 
In Fig.~\ref{fig_fom}, we also show the expected constraining power from the Simons Observatory for two configurations: SO-Baseline (blue dash-dotted) and SO-Goal (orange dash-dotted), based on the noise levels in \citet{simonsobservatorycollab19}.
\refresponse{
%We use the original SO noise model code\footnote{\url{https://github.com/simonsobs/so_noise_models}} to get the noise curves but use the same multipole ranges described in \S~\ref{subsec_fisher_formalism} for forecasting. 
We use the original SO noise model code\footnote{\url{https://github.com/simonsobs/so_noise_models}} to get the noise curves for the individual frequency bands and combine information from all the bands using the LC technique described in \S\ref{sec_ilc}. 
We use the same multipole ranges described in \S~\ref{subsec_fisher_formalism} for forecasting constraints from SO. 
}
As a check, we compare the constraints that we get from our forecasts for SO against what is reported in \citet{simonsobservatorycollab19} but note that we do not include \planck{} unlike SO's original forecasts. 
%These are provided in Table~\ref{tab_so_forecasts}. 

Comparing our forecasts for SO to Table 3 of \citet{simonsobservatorycollab19}, our forecasts are slightly better for $H_{0}$ and $\omch$ but they agree within 10\%. 
For $n_{s}$ and $\ombh$, we obtain 15-35\% worse constraints, which is presumably because we do not add the large-scale information from \planck. 
On the other hand, we obtain roughly 35\% (20\%) better constraints on $A_{s}$ ($\taure$). 
When we increase the prior uncertainty on optical depth to $\sigma(\taure) = 0.009$ to match SO's forecasting code (see caption of Table 3 in \citealt{simonsobservatorycollab19}), our results on $A_{s}$  and $\taure$ match the numbers quoted in \citet{simonsobservatorycollab19}.
%We attribute these differences to the exclusion of \planck{} data and also due to different multipole ranges used for forecasting. 
Switching to SO-Goal configuration reduces the errors on all parameters by 5\% except for $\omch$ where the constraining power improves by 15\%.

We also note from Fig.~\ref{fig_fom}, the constraining power of the SPT-3G \allspt{} survey for \llcdm\ parameters will be roughly similar to the first two years of SO-Baseline. 
Such similar levels of constraining power between SPT and SO along with significant sky overlap (see Fig.~\ref{fig_spt_footprints}) will allow for comparisons of the two datasets at both the power spectrum and map level (such as the ACT-\planck\ and SPT-\planck\ comparisons shown in \citealt{louis14} and \citealt{hou18}) which will provide useful opportunities for identifying and/or limiting systematic errors in individual datasets.

\subsubsection{Real-world effects}
\label{sec_realworldeffects}
In this section, we examine how our forecasts change as we vary our baseline modeling assumptions, and introduce additional non-ideal aspects into our model of the data on which the forecasts are based. 
We modify our noise curves to account for filtering effects and to include a higher level of complexity in the noise due to the correlated atmospheric signal on large scales in the temperature data. 
We also investigate the degradation in constraints that could come from galactic foregrounds. 
We also check the impact of including uncertainties in the calibration of our temperature and polarization maps. 
As detailed below, we find the degradation of parameter errors relative to the baseline case, from all of these effects, is $\lesssim 10\%$.
Finally, we build further confidence in our forecasts by comparing them to constraints from an ongoing MUSE analysis of the two-year \mainfield{} data.

%However, in all the cases we find that the forecasts are robust to these effects and departures from the baseline are less than 10\%. 

%\paragraph{Atmospheric noise}
\specialparaheader{Atmospheric noise}
The first thing we examine is the impact of the filtering applied to the detector time-ordered data (TOD) before it is processed into maps. 
To avoid excess noise in the scan direction in the resulting maps, primarily as a result of the large-scale atmospheric noise, we apply filters to remove the long-timescale information from the detector TOD. 
While this improves the quality of the maps, it results in a reduction of data volume that increases the final power spectrum errors by a factor equivalent to raising map noise levels by roughly 10\%. 
We test the impact of this by increasing the white noise levels in Table~\ref{tab_survey_specs} by 10\%. 
We found that this results in changes to the parameter constraints at the $\leq 1\%$ level.
%The first thing we examine is the impact of filtering out atmospheric noise.
%Before converting the raw data to maps, we employ filtering techniques on the timestreams to remove undesired signals like the excess noise along the scan direction. 
%While these improve the quality of the CMB maps, these result in a reduction of data volume and subsequently enhances the noise levels of the maps, typically by 10-20\%. 
%We test the impact of this by increasing the white noise levels in Table~\ref{tab_survey_specs} by 10-20\%. We found that this results in negligible changes to the parameter constraints at $\leq 1\%$ level.
Secondly, since SPT-3G uses multichroic detectors that can simultaneously observe in multiple frequencies, the large-scale atmospheric noise is expected to be correlated between bands. 
From the 2019-20 dataset, we estimate the correlation at the largest scales to be $\sim 90\%$ between the 150 and 220 GHz bands, and around $70\%$ between the 90 and 150 GHz bands as well as between the 90 and 220 GHz bands. This degree of correlation diminishes as we move to small scales. 
To take this into account, we set the cross-frequency noise to be correlated across bands and rerun the LC step to get the residuals. 
This correlated noise, in fact, reduces the large-scale LC residuals slightly. 
When we propagate these to parameter constraints, we find changes to the parameter constraints at the $\leq 1\%$ level,
%When we propagate these to parameter constraints, we find only marginal changes to the parameter constraints at $\leq 1\%$ level 
indicating that our forecasts are robust to these details about the noise spectra.

%Filtering techniques are employed to improve the quality of the CMB data by removing or reducing  unwanted signals from the raw data. These can lead to enhancement of noise levels and impact the constraining power. We enhance the white noise levels by 10-20\% and test the impact on the forecasts. We also account for correlated atmospheric noise between frequencies by measuring the correlation coefficients from our data. At the largest scales we find roughly 90\% correlation between 150 GHz and 220 GHz bands, while the correlation coefficient for 90 GHz and 150GHz, and 90 GHz and 220 GHz is around 70\%. Both these tests, lead to $\leq 1\%$ shift in the parameter constraints.

%\paragraph{Galactic foregrounds and mask}
\specialparaheader{Galactic foregrounds and mask}
Given that the \extfield{} survey extends close to the plane of the galaxy (see Fig.~\ref{fig_spt_footprints}), it is crucial to understand the impact of galactic foregrounds on the parameter constraints. 
To assess this, we make use of the PySM simulations \citep{thorne17} of galactic foregrounds and the galactic foreground masks (\texttt{GAL070, GAL080, GAL090}) produced by the \planck{} collaboration \citep{planck15-13}. 
The names of the above masks correspond to the sky retained after masking the galactic plane. 
For instance, \texttt{GAL090} masks 10\% of the region with a high level of foregrounds close to the galactic plane while \texttt{GAL070} removes 30\% of the sky and is the most conservative amongst the three. 

We use these masks and calculate the level of galactic dust and synchrotron signals in PySM simulations in the following regions:
Case (A) \texttt{GAL080-GAL070}; Case (B) \texttt{GAL090-GAL070}; and Case (C) \texttt{GAL090-GAL080}. 
The three cases correspond to the differences between the respective masks; for instance,
\texttt{GAL080-GAL070}
represents the sky area included in mask \texttt{GAL080} but excluded in \texttt{GAL070}.
As expected, we find the galactic foreground in Case (C) to be the highest followed by Cases (B) and (A). 

We perform a conservative test by assuming that the galactic foreground levels across the entire \extfield{} survey match one of these three cases. With this assumption, we derive the LC residuals and the subsequent impact on the parameter constraints. 
Even for Case (C), where the galactic foregrounds are the highest, we find the constraints degrade by only $8\%$. 
This is not surprising given that the galactic foregrounds are mostly important on the largest scales and the constraining power of SPT-3G is predominately driven by the high-$\ell$ region. 

In the preceding sections, we mentioned that the observing strategy to minimize galactic foreground contamination in the \extfield{} involves observing within the confines of the \planck{} \texttt{GAL080} mask, which corresponds to roughly $6000\,\sqdeg$ of sky coverage. In the scenario where we might need to adopt a more conservative \planck{} \texttt{GAL070} mask, we found that this would lead to a modest degradation in parameter constraints, at 4-5\%. While we do not address potential biases due to mismodeling of the galactic foregrounds, we remain confident that any such biases would be minimal, given that the levels of galactic foregrounds are significantly lower than the CMB power spectra levels at the multipoles relevant to this analysis.

%Galactic foregrounds contribute to additional variance in the ILC noise curves depending on the assumed foreground levels. This is particularly significant for the \extfield{} survey given its proximity to the galactic plane. To assess the impact we considered different levels of foreground by applying several \planck{} masks to the observing regions of the \extfield{}: GAL70, GAL80, GAL90 as well as differential areas between these masks specifically, GAL80-GAL70, GAL90-GAL70 and GAL90-GAL80 arranged in the order of increasing foreground levels. We found that even for the most conservative case (GAL90 - GAL80) the degradation in the constraints was $\leq 8\%$.

%\paragraph{Temperature and polarization calibration}
\specialparaheader{Temperature and polarization calibration}
Next, we assess the impact of calibration uncertainty on the temperature and polarization maps by including two absolute calibration parameters, $T_{\rm cal}$ and $P_{\rm cal}$ \citep[as defined in][]{dutcher21}, in our forecasts and marginalizing over them. 
%$T_{\rm cal}$ and $P_{\rm cal}$ are scaling factors applied to the maps, such that $TT \propto T_{\rm cal}^2$, $TE \propto T_{\rm cal}P_{\rm cal}$, and $EE \propto P_{\rm cal}^2$.
In the case where we do not apply a prior, we find a significant $\times2$ hit on some of the parameters, which is not surprising since parameters like $A_{s}$ are fully degenerate with the calibration parameters. 
By comparing SPT-3G Main survey maps from 2019-2020 to \planck\ maps in that same region, we are able to constrain $T_{\rm cal}$ and $P_{\rm cal}$ to roughly 0.2\% and 1\%, respectively (where the errors are a combination of statistical uncertainty in the comparison and uncertainties in the \planck\ calibration). If we adopt these as priors, the resulting degradation in the parameter constraints are $\le 5\%$ for all parameters. If we loosen the $T_{\rm cal}$ prior to 1\%, the degradation increases to $\sim 20\%$. If we assume either parameter is estimated perfectly and allow for uncertainty in only one calibration parameter, we find the resulting degradation to be negligible, even without any prior, consistent with the results from \citet{galli21}.

%\kp{Work in progress}
%Finally, we compare the constraints from our forecasts to those from the MUSE analysis of the first two year data from \mainfield{} survey. This analysis includes realistic effects like anisotropic filtering, noise derived from sign-flip maps, masking etc. which are not accounted for in our forecasts. We find that the degradation in the parameter constraints from these real-world effects in $\leq 10\%$. 
Finally, we replace the covariance matrix for the 2-year \mainfield{} survey from this work with the one from an ongoing MUSE analysis performed on the real data from the same survey. The latter takes into account certain effects that are present in the real data but not taken into account in our forecasts. These include anisotropic noise, anisotropic filtering, and masking. 
With this covariance swap, we find that the degradation in the parameter constraints is $\leq 10\%$.

%Finally, we assess the impact of miscalibration by including the parameters, $Tcal$ and $Pcal$ in our forecasts and marginalizing over them. Since we calibrate our power spectra measurements against \planck{} data, we apply tight priors of 0.2\% and 1\% on these parameters respectively. We find that the resulting degradation in the parameter constraints are less than 5\%.

\subsection{Single-parameter extensions to \llcdm\ }
\label{sec_single_parameter_extensions}

\begin{figure}
    \centering
    \ifdefined\apjformat
        \includegraphics[width=0.48\textwidth, keepaspectratio]{single_param_ext.pdf}
    \else
        \includegraphics[width=0.48\textwidth, keepaspectratio]{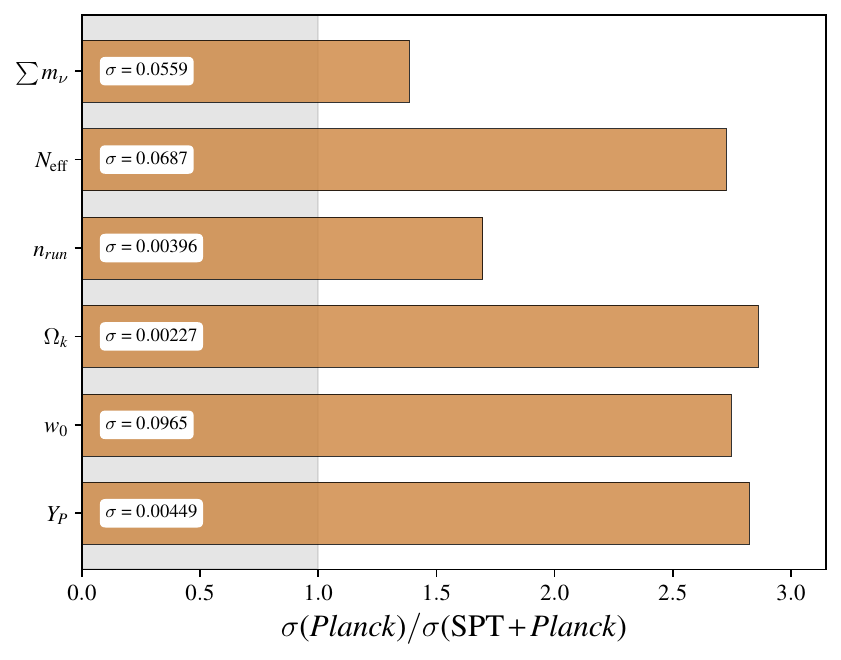}
    \fi
    \caption{Ratio of 1$\sigma$ constraints on single-parameter extensions from \planck{} to those forecasted from the combination of \allspt{} and \planck{}. 
    The gray shaded region corresponds to the current best constraints from \planck{} alone. 
    The actual errors are expressed in the text boxes.
    }
\label{fig_single_param_constraints}
\end{figure}

\begin{figure}
    \centering
    \ifdefined\apjformat
        \includegraphics[width=0.48\textwidth, keepaspectratio]{lcdm_degradation_due_to_single_parameter_extension.pdf}
    \else
        \includegraphics[width=0.48\textwidth, keepaspectratio]{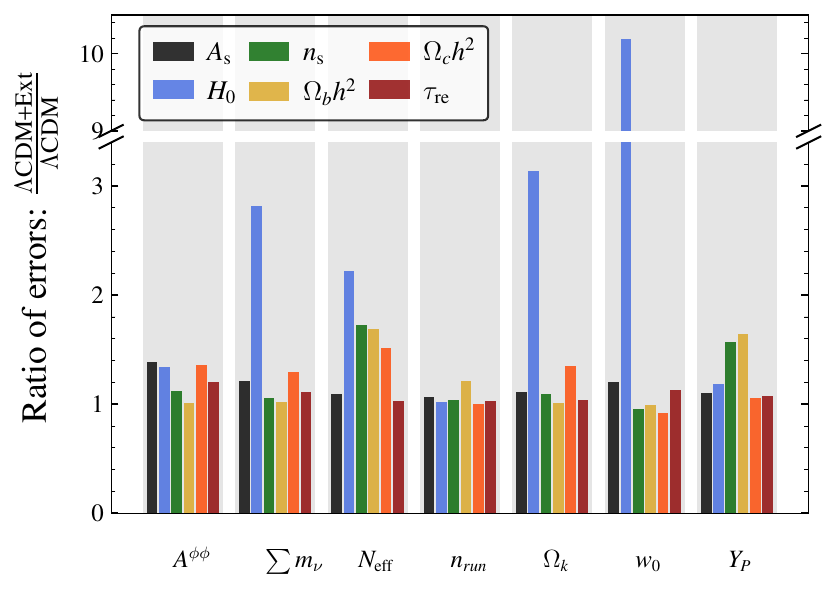}
    \fi
    \caption{Degradation in the \llcdm\ constraints due to single-parameter extensions compared to the original constraints reported in Fig.~\ref{fig_lcdm_constraints_spt_vs_plank} for the \allspt+\planck{} dataset. The horizontal axis shows the parameter that is varied in our extensions. Except for $H_{0}$, which is affected by a geometric parameter degeneracy, the degradation in constraints for other \llcdm\ parameters is quite mild. }    
\label{fig_lcdm_degration_due_to_single_param_ext}
\end{figure}

Now we turn our attention towards extensions to the \llcdm\ model. 
For extensions, we always combine \sptnew{} with \planck, and use the analytic method.
We consider 7 different cases by varying the following parameters one at a time along with the 6 \llcdm\ parameters: 
dark-energy equation of state ($\wde$), Helium mass fraction ($\yp$), running of the spectral index ($\nrun \equiv \nrunexplicit$), 
sum of neutrino masses ($\summnu$), effective number of neutrino species ($\neff$) and the spatial curvature parameter ($\omk$). 
The forecasted constraints for \allspt{} are reported in Fig.~\ref{fig_single_param_constraints}. 
As before, we show the improvement in constraints compared to \planck{} using the bars. The actual parameter constraints are given in the text boxes. 
The constraints from other \sptnew{} surveys are presented in Table~\ref{tab_ext_param_table} of Appendix~\ref{appendix_cosmo_constraints}. 

We find a significant ($\times2 - \times3$) reduction of the errors in many of these parameters, namely $\omk$, $\wde$, $\yp$, and $\neff$. 
For other parameters, namely $\summnu$, and $\nrun$, the improvement is $\times1.4-\times1.7$ compared to \planck. 
Some of these improvements are driven by the inclusion of lensing ($\omk, \wde$) while others come from higher-redshift information in the damping tail of the CMB, whose extraction benefits from wide sky coverage ($\nrun$, $\yp$, and $\neff$). 

We also investigate the degradation in \llcdm\ constraints due to degeneracies with the extension parameters.
We present these error degradations in Fig.~\ref{fig_lcdm_degration_due_to_single_param_ext} for \allspt+\planck{}; the full list can be found in Table~\ref{tab_lcmd_degradation_due_to_ext} of Appendix~\ref{appendix_cosmo_constraints}. 
As can be inferred from the figure, the constraints on the Hubble constant $H_{0}$ degrade significantly when we free $\theta \in [\summnu, \neff, \omk, \wde]$. Note that these degradations largely disappear with the addition of external datasets that constrain the redshift-distance relationship down to $z \lesssim 0.1$, namely  supernova brightness measurements. This is the reason why these extensions are not viable solutions to the $H_0$ tension \citep{knox20}.
The degradation is less severe ($\sim \times 1.6$) for $n_{s}, \ombh$ when we vary $\theta \in [\neff, \yp]$. 
For the rest, the constraints are robust and the degradation is 
$\lesssim \times1.3$. 

In expanding the parameter space from \llcdm\ to \llcdm+$w_0$, we notice an unexpected tightening in the constraints for parameters $\ombh$, $\omch$, and $n_s$. We suspect that this arises from our use of a Gaussian approximation for a posterior that is significantly non-Gaussian.  The non-Gaussianity is visually evident in the one and two-dimensional marginal posteriors derived from the appropriate \planck{} chains. With the addition of an $H_0$ prior, the range of acceptable $w$ values tightens significantly, reducing the degree of non-Gaussianity, and this anomalous behavior of our forecasts no longer appears.

\subsection{Double-parameter extensions to \texorpdfstring{\llcdm\ }{LCDM}}
\label{sec_double_parameter_extensions}

\begin{figure*}
    \centering
    \ifdefined\apjformat
        \includegraphics[width=\textwidth, keepaspectratio]{double_param_ext_mnu_nnu_yhe_nrun.pdf}
    \else
        \includegraphics[width=\textwidth, keepaspectratio]{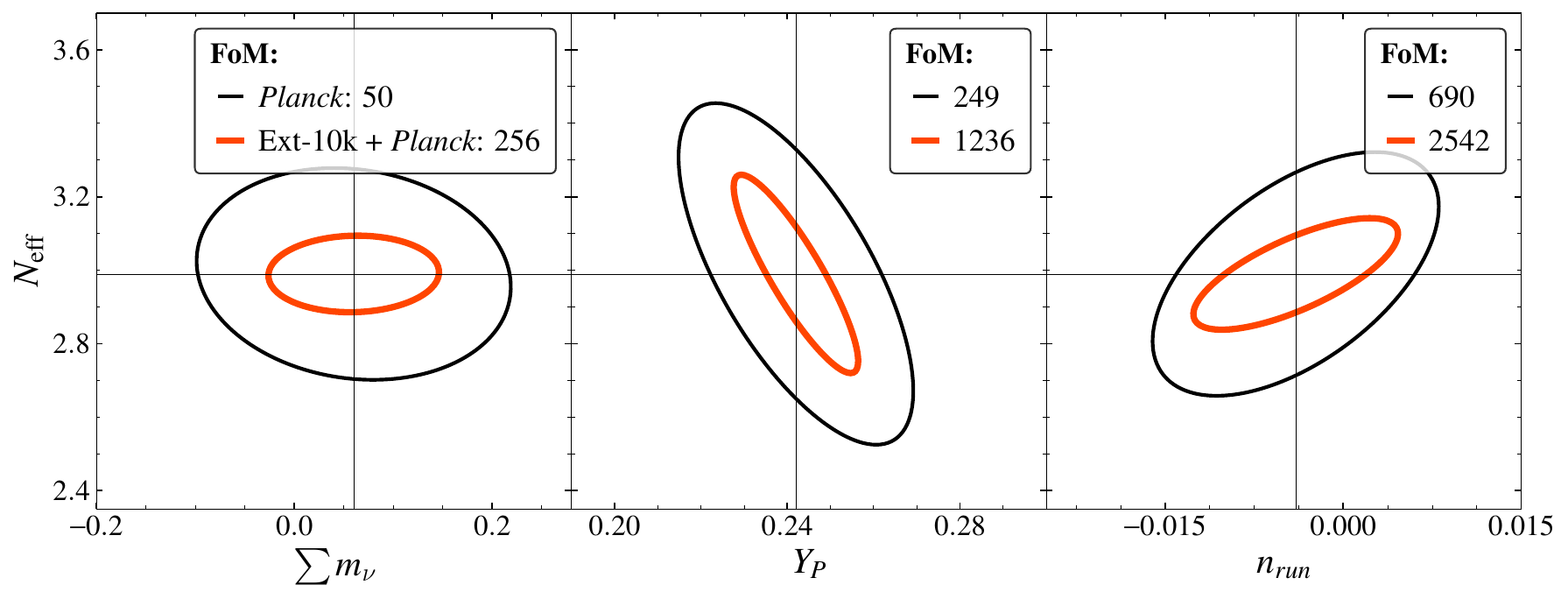}
    \fi
    \caption{Marginalized two-dimensional parameter contours for the three pairs of double-parameter extensions to \llcdm\ considered in this work. The red contours represent the joint constraints from the \allspt{}+\planck, and the black contours show the constraints from \planck{} alone. The inverse of the allowed volume of the two-dimensional parameter space (FoM) are given in the legend. Compared to the constraints from \planck{} alone, we find reductions in individual parameter errors that approach or exceed a factor of two except for $\nrun$, where we obtain $\times 1.3$ reduction. 
    }
    \label{fig_two_param_ext}
\end{figure*}

We also explore two-parameter extensions similar to those considered in \citet{planck18-6}.
%We also explored a few 2-parameter extensions similar to those considered in \citet{planck18-6}. 
These are (a) $\neff$ and $\yp$, (b) $\neff$ and $\nrun$, and (c) $\neff$ and $\summnu$. 
In Fig.~\ref{fig_two_param_ext}, we present the marginalized two-dimensional contours for the above extensions from both \planck{} (black) and \allspt+\planck{} (orange). 
%The constraints on each parameter are given in the legend. 
The FoM for the two-parameter extensions, calculated similar to Fig.~\ref{fig_fom}, is given in the legend. 
In most cases, we find an improvement of greater than a factor of 2 relative to \planck{} alone. 

In case (a), we vary the effective number of neutrino species $\neff$ and the fraction of baryonic mass in helium $\yp$  simultaneously.
Such constraints are useful for constraining scenarios in which the value of $N_{\rm eff}$  at big bang nucleosynthesis differs from that at recombination \citep{millea15, cyburt16, zyla20}.
Since both these parameters affect the damping tail of the CMB, they are partially degenerate and, as expected, the constraints on both parameters degrade by more than $\times 2$ compared to the single-parameter extensions in Fig.~\ref{fig_single_param_constraints}. 

In case (b), we vary $\neff$ and $\nrun$ which are also degenerate to some level leading to $\times 1.5$ weaker constraints compared to Fig.~\ref{fig_single_param_constraints}. 

In case (c), we vary $\neff$ and the sum of neutrino masses $\summnu$ since one can expect some level of correlation between the two \citep{planck18-6}. 
However, this correlation depends on the datasets being considered. 
In our case, $\summnu$ is primarily constrained by lensing while $\neff$ constraints are dominated by the small-scale $TT$, $EE$, and $TE$ power spectra. 
Subsequently, the degradation in the constraints on these two parameters is negligible compared to what we obtain for the single-parameter extensions in Fig.~\ref{fig_single_param_constraints}. 

\section{Power spectrum predictions from \llcdm\ and alternatives}
\label{sec_lcdm_alternatives}

\begin{figure*}[]
\centering
\ifdefined\apjformat
    \includegraphics[keepaspectratio]{BandpowerForecast_new.pdf}
\else
    \includegraphics[keepaspectratio]{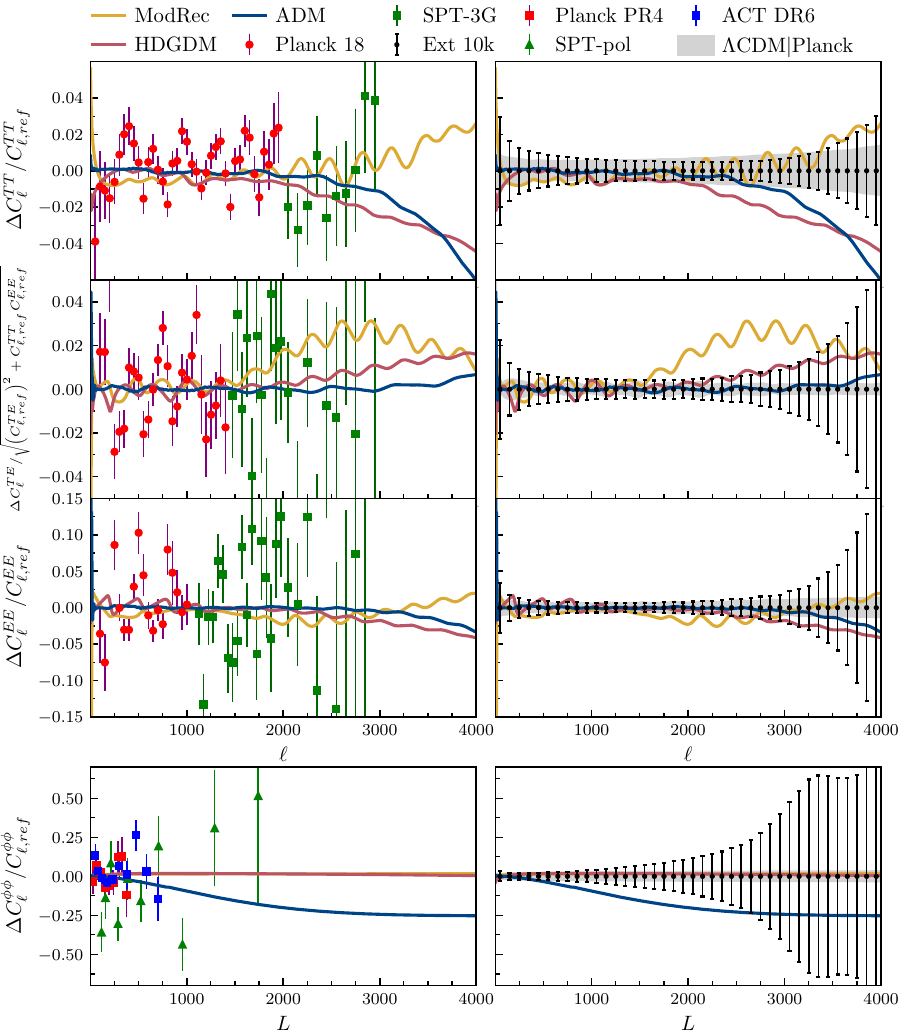}
\fi
\caption{\textit{Left panels:} Current data from CMB experiments and some alternative models with either higher $H_0$ or lower $\sigma_{8}$ (see text). Except for lensing data, only the most constraining observations in a given multipole range have been plotted for clarity. Each spectrum has the best-fit \llcdm\ spectrum given \planck{} subtracted as a reference and is normalized in an $\ell$-dependent manner to reduce dynamic range, as described in the $y$-axis label. Note that for the ADM model, neither the model prediction nor the reference model includes non-linear corrections to the lensing power spectrum, as existing non-linear correction procedures do not account for the modified dark sector physics. All spectra are taken from chains constrained using \planck{} and BAO data, but the specific BAO datasets used differ between models. The HDGDM constraints also used uncalibrated supernovae data as well as the SH0ES $H_0$ measurement. \textit{Right panels:} The same models as in the left panels, forecasted bandpower errors for \allspt, and the 68\% confidence region given \planck\ and the assumption of \llcdm. The bandpower errors have been binned with $\Delta \ell = 100$. Note that the \llcdm\ predictions conditioned on \planck{} are very tight, even in regions where \planck{} data on their own are completely unconstraining.  
}
\label{fig_alternative_models}
\end{figure*}

In this section we discuss, in a qualitative manner, prospects for using \sptnew{} data to detect departures from \llcdm\ predictions for CMB power spectra conditioned on \planck{} data. 
Many alternative models that resolve cosmological tensions, such as the $H_0$ and $\sigma_8$ tensions, predict power spectra that differ from \llcdm\ predictions at small angular scales, and we use a selection of such models to guide our discussion. 
In Fig.~\ref{fig_alternative_models}, we present the \llcdm\ predictions for angular power spectra given \planck{} data, the predictions of alternative models that fit the \planck{} data well while either having a high $H_0$ or a low $\sigma_8$, and the forecasted \allspt{} error bars. 

%the differences between power spectra predictions of \llcdm\ and some alternative models that can reconcile the two main cosmological tensions, all in h show the highly precise power spectra predictions that follow from assuming the \lcdm\ model conditioned on Planck data, the power spectra for a few \lcdm\ extensions that can reduce cosmic tensions, and the forecasted SPT-3G power spectrum error bars. These are all shwon in Fig.~\ref{fig_alternative_models}. 
%We compare those predictions to the forecasted SPT-3G error bars and examine the extent to which \sptnew{} can differentiate between \llcdm\ and alternative models which have been proposed as solutions to existing cosmological tensions \srini{Citations.} \citep{xxx}. In doing so, we demonstrate the potential for \sptnew{} to probe the physics of the baryon-photon plasma of the early universe, as well as the potential to provide insight into hitherto \srini{\sout{unprobed} unexplored} dark sector physics. 
The selected alternative models are (a) a phenomenological model of non-perturbatively modified recombination (ModRec) \citep{Lynch2024}; (b) an atomic dark matter (ADM) model arising from a hidden dark sector charged under a U(1) gauge interaction \citep{hughes23}; and (c) a model featuring an additional dark fluid with an equation of state parameter that is a free function of redshift, hereafter ``high dimensional generalized dark matter" (HDGDM) \citep{meiers23}. 

We present one set of spectra from each of these models in Fig.~\ref{fig_alternative_models} as a relative shift $\Delta C_{\ell}^{X} = C_{\ell}^{X} - C_{\ell, {\rm ref}}^{X}$ compared to the reference \llcdm\ mean \planck{}  $C_{\ell, {\rm ref}}^{X}$, normalized as described in the figure caption. Here $X$ corresponds to one of the primary lensed CMB $TT/EE/TE$ or the lensing $\phi\phi$ power spectra. Each of these spectra are drawn from posterior distributions conditioned on \planck{} and BAO data, although the HDGDM spectra also includes uncalibrated supernovae data in the fit. It should be noted that the specific BAO datasets used in these constraints differ: please see the appropriate references for details.

The left-hand panels of Fig.~\ref{fig_alternative_models} show these spectra in relation to existing CMB measurements from \planck{}, \sptpol{} \citep{SPT:2019fqo}, \sptnew{} 2018 \citep{SPT-3G:2021eoc} and ACT DR6 \citep{qu24}. 
For angular scales that have been probed by multiple experiments, only the most constraining dataset has been plotted. This set of panels highlights that although these models make predictions that differ from \llcdm{}, they are currently not ruled out by the data. To the right, we show the 68\% confidence region for angular power spectrum predictions conditioned on \planck{} data and assuming \llcdm{}, along with forecasted bandpower errors for \allspt{} in black.

The \planck{} data, along with the assumption of \llcdm{}, make tight predictions for the power spectrum residuals even at angular scales not probed by \planck{}. This is evident from the righthand panels of Figure~\ref{fig_alternative_models}, where the gray confidence interval extends to no more than about 2\% for the $TT$, $TE$, and $EE$ power spectra. This range represents the expected region that the power spectra observed by \sptnew{} will fall in if measurements continue to be consistent with \llcdm{}. As is evident in these panels, the alternative models presented here exhibit deviations which are larger than this expected range, and larger than the expected precision of \sptnew{}. We now discuss each of these models in turn.

%The ModRec and HDGDM models are examples of a general class of models which address the the Hubble tension by introducing new physics in the two decades of scale factor preceding recombination, an epoch which has been identified as promising for model builders (\cite{Knox:2019rjx}). 

The potential role of changes to recombination in alleviating the $H_0$ tension has been explored in \citet{chiang18, hart19, sekiguchi20, lee22, cyrracine21}. Physical models which have a modified recombination epoch include models with varying fundamental constants \citep{Planck:2014ylh, Hart:2017ndk, hart20, Sekiguchi:2020teg}, primordial magnetic fields \citep{Jedamzik2020Relieving, Galli:2021mxk, Rashkovetskyi:2021rwg}, or energy injection from decaying dark matter \citep{Galli:2013dna, Slatyer:2016qyl, Finkbeiner:2011dx}. The ModRec model presented here is an extension of \llcdm\ that allows for a purely phenomenological modification of the ionization fraction as a function of redshift through the epoch of hydrogen recombination, in order to study the role of these changes independently of the details of particular physical models. We note that the different modified ionization fractions predicted by the above physical models can be well approximated within the ModRec model using seven parameters characterizing deviations from the standard recombination scenario.  

This additional freedom allows for higher values of $H_0$ to provide good fits to the \planck\ data. The ModRec model accommodates a high value of $H_0$ by altering the recombination process, shifting last scattering to higher redshifts and increasing the width of the visibility function. This reduces the comoving size of the sound horizon at recombination, so that the shorter distance to last-scattering associated with a higher $H_0$ does not shift the well-determined angular size of the sound horizon. It also impacts the damping scale, an effect ameliorated by the change to the visibility function width.

In Fig.~\ref{fig_alternative_models} we show predictions within the ModRec model for a set of parameters which fit \planck{} and eBOSS BAO data at least as well as a \llcdm\ best-fit, while having \mbox{$H_0=70.48$ km/s/Mpc}. To maintain the fit with \planck{} data, residuals with respect to the fiducial cosmology get shifted to higher multipoles, which, while unprobed by \planck{}, are expected to be measured with high SNR by \sptnew{}. We see that \sptnew{} has clear potential to rule out such a model, particularly through measurements of the $TE$ spectrum. We note that without including BAO data, much higher values of $H_0$ are achievable within the ModRec model while staying consistent with \planck{}. In this case, if \sptnew{} measurements remain consistent with \llcdm{}, solutions to the $H_0$ tension which modify recombination will be seriously challenged by CMB data alone. 
 
Another approach to resolving the $H_0$ tension is to leave the recombination history unmodified and to instead introduce more freedom in the dark sector. The HDGDM model is another phenomenological model that uses the generalized dark matter (GDM) model of \citet{hu98} and allows the equation of state parameter for the GDM to be an arbitrary function of redshift. \citet{meiers23} used such a model to approximate, and perturb around, a more physical model, the ``step-like" model of \citet{aloni22}. 
These authors 
introduce a massive particle which becomes non-relativistic before decaying, in equilibrium, into a massless final state. 
The equation of state parameter $w(z)$ for such a species begins at 1/3, drops as the particle becomes non-relativistic, and returns to 1/3 in the final massless state. 
This new component affects the expansion history in a non-trivial way, and alters the time-dependent gravitational driving of the plasma. They found that a transition at $z \sim 20000$ could accommodate a higher $H_0$. 
In Figure~\ref{fig_alternative_models}, we present the best-fit spectrum of the ``step-like restricted" model of \citet{meiers23} constrained using the ``$\mathcal{D}+$" data combination in that work, which includes \planck\, BAO, uncalibrated supernovae data, as well a Gaussian likelihood centered at the SH0ES measurement \mbox{$H_0 = 73.04 \pm 1.04$ km/s/Mpc} \citep{riess22}. This model has \mbox{$H_0 = 71.88$ km/s/Mpc}, illustrating how the HDGDM model can accommodate higher values of $H_0$. However, the TT damping tail predictions from this model differ from the \llcdm\ prediction using the \planck\ mean parameters at a more than 2\% for $\ell>3000$, which should be distinguishable with \sptnew.

In addition to the $H_0$ tension, in recent years there has been increased attention to the potential discrepancy between different measurements of the amplitude of matter fluctuations on \mbox{$8h^{-1}$ Mpc} scales, $\sigma_8$. 
Inferences from \planck{}, that depend on assuming the \llcdm\ model,  give a value of $\sigma_8 = 0.811 \pm 0.006$, whereas the KiDS-1000 survey, measuring cosmic shear, gives a value of $\sigma_8 = 0.76^{+0.025}_{-0.020}$ \citep{planck18-6, heymans20}. Although, DES sees less of a $\sigma_8$ tension compared to KiDS \citep{abbott23c}.
In the ADM model considered here, this tension is addressed through the introduction of two dark fermions interacting with a dark photon, where the dark fermions make up some fraction $f_{\text{ADM}}$ of the dark matter. This results in dark sector dynamics analogous to the dynamics of the baryon-photon plasma. The pressure support from the dark photon at early times, prior to the formation of dark atoms, suppresses the growth of structure. 
In general, this leads to effects in the CMB power spectra observable by \planck{} but if the dark sector temperature is cool enough, the epoch of significant dark photon pressure support is early enough, and the \planck{} observables are relatively unaffected. More direct inferences of $\sigma_8$ are impacted by smaller scales than those to which \planck{} is sensitive; so, at the right dark photon temperature with a suppressed $\sigma_8$, \planck{} observables are unaffected. In Fig.~\ref{fig_alternative_models}, we present the best-fit model identified in \citet{hughes23}, coming from a region of parameter space consistent with \planck{} data, with  $\sigma_8 = 0.7869$, a low dark photon temperature, and high $f_{\text{ADM}}$. For this fit, both CMB data from \planck{} were used, as were BAO measurements (see \citealt{hughes23} for details). %In this case, the main observable effect of high $f_{ADM}$ is a reduction in lensing power, related to the suppressed matter power spectrum. 

%\kp{Skip: The depth of \sptnew{} in both the Main-1500 and Ext-10k configurations is expected to provide tight constraints on lensing power, as is evident from the lower panel of Figure \ref{fig_alternative_models}. If the $\sigma_8$ tension is a result of a cool dark sector, we expect lensing measurements from \sptnew{} to be able to detect this with high confidence. }

%\kp{Move to conclusions: From this survey of alternatives to \llcdm\, we conclude that measurements of temperature, polarization, and lensing power spectra from \sptnew{} promise to clarify existing cosmological tensions. These measurements stand to further improve our understanding of the primordial plasma and early thermal history of the universe, as well as to shed light on the myriad possible forms that dark matter may take on. }

%Figure \ref{fig_alternative_models} presents predictions from these three models along with forecasted errors from the \allspt{} configuration. For each model, we expect \sptnew{} to be able to distinguish between these alternatives and \llcdm\. 

From this survey of alternatives to \llcdm\, we conclude that measurements of temperature, polarization, and lensing power spectra from \sptnew{} are capable of clarifying existing cosmological tensions. 
%These measurements stand to further improve our understanding of the primordial plasma and early thermal history of the universe, as well as to shed light on the myriad possible forms that dark matter may take on. \srini{May be move this to conclusion.}

%\clearpage
%--------------------------------------------------------------------------%

\section{Conclusions}
\label{sec_conclusion}

We presented forecasts of constraints on cosmological parameters that can be derived from the estimates of lensing, temperature and E-mode polarization power spectra to come from completed and ongoing \sptnew{} sky surveys.
%to come from the ongoing \sptnew{} sky surveys. 
We made forecasts for three combinations of the three \sptnew{} surveys of varying sizes and depths, with particular emphasis on the combination of all three of these surveys, called \allspt{}, 
covering $\sim 10000\,\sqdeg$ (25\% of the sky).
%survey which is a: the \mainfield{}\ survey spans an area of $1500\sqdeg$, \mainplussummerfield{} encompasses \mainfield{} and \summerfield{} and spans $4000\sqdeg$ and \allspt{} that encompasses \mainfield{}, \summerfield{}, and \extfield{} surveys summing up to a total coverage of $10000 \sqdeg$, or approximately 25\% of the sky.
%We forecasted the precision with which we can measure the temperature and E-mode polarization power spectra, their cross power spectrum, and the power spectrum of the gravitational lensing potential.
We showed that \sptnew{} will enable some powerful tests of the \llcdm\ model. 

We showed that the constraining power of the full \sptnew{} dataset will surpass
%We expect that the constraining power of the full \sptnew{} dataset will surpass 
that of \planck\ for the power spectra at modes $\ell \gtrsim 1800$ in $TT$, $\ell \gtrsim 800$ in $TE$, at $\ell \gtrsim 450$ for $EE$ and at $L \gtrsim 30$ in $\phi \phi$.  We calculated the predictions of the \llcdm\ model, conditioned on \planck{} data, and found that they are highly precise in the power spectra on these scales, setting us up for a powerful test of \llcdm.

We also demonstrated that \sptnew{} will facilitate testing of the \llcdm\ model at the level of cosmological parameters. 
By propagating the uncertainties on power spectra to uncertainties on cosmological parameters, our analysis suggests that data from \sptnew{} will be able to achieve similar, or in some cases, tighter constraints than those from \planck. Since these constraints incorporate higher weights from polarization, from small angular scales in temperature, and from lensing when compared to those from \planck, a consistency check between the two will be a powerful test of \llcdm. 
The constraints on the current expansion rate ($H_0$) and baryon density ($\ombh$) improve by roughly a factor of 2 when compared with the constraints from \planck, while for the dark matter density ($\omch$), the improvement is a factor of 1.5.
We find that the full \sptnew{} dataset is expected to achieve an improvement in the \llcdm\ FoM of 130; i.e., \sptnew{} will reduce the allowed six-dimensional parameter volume by a factor of 130 compared to \planck.
%We find that the full \sptnew{} dataset is expected to achieve an FoM of 130; i.e., reduce the allowed six-dimensional parameter volume by a factor of 130 compared to \planck.

 This reduction in volume is comparable to what is expected from the first two years of the SO-Baseline dataset. This similarity in constraining power, plus the substantial amount of overlap in sky coverage, provides an opportunity for significant map-level consistency tests. These can be used to detect, or limit, sources of systematic error, increasing the robustness of the parameter determinations from both SO and SPT-3G. 
 
%Furthermore, if the \llcdm\ model is validated through this comparison, it demonstrates that \sptnew{} will significantly enhance the precision of these parameter estimates. 
%and found that the data from the first two years of the \mainfield{} survey, anticipated to be released later this year, will offer constraints that are comparable to those currently available from \planck.
%We demonstrated that \allspt{} will improve the six-dimensional parameter volume by $\times 130$ compared to \planck. 
%A majority of this improvement is enabled by data from the 6000 $\sqdeg$ \extfield{} field, observations of which are currently underway and will continue through the 2024 austral winter. 
%We found that the constraints from the first two years of the \mainfield{}, expected to be released in the upcoming year, will surpass those currently available from \planck.
%We also found that the reduction in the allowed six-dimensional parameter volume given \sptnew{} is roughly comparable to that expected from the first two years of the SO-Baseline dataset. 
%The excellent overlap in the sky coverage between \sptnew{} and SO-Baseline will also allow for cross-correlation analysis including map-level comparisons that will be free from systematics, if any, that may affect results from each dataset individually. 

We also explored how well we can test \llcdm\ by forecasting constraints on single- and double- parameter extensions  with the combination of \planck+\sptnew. We found that the combined data will result in considerable improvements over \planck\ alone.
For example, we found that the constraints on single-parameter extensions improve by more than $\times 2$ for $\theta \in [\omk, \neff, \wde, \yp]$ and roughly $\times 1.5$ for $\theta \in [\nrun, \summnu]$ compared to \planck-only constraints. 
We also showed that including these extensions does not significantly degrade the constraints on the \llcdm\ model parameters, except for $H_{0}$, which is highly degenerate with some of the extensions like $\wde$. 
 
In addition to showing the highly-precise predictions of the \llcdm\ model on angular scales to be measured by \sptnew{}, we showed that there are also alternative models that are consistent with current data, with significantly different predictions on these same angular scales.
%\sptnew{} will also be capable of testing alternatives to the \llcdm\ model that make different predictions for the power spectra at scales to which \sptnew is particularly sensitive.  
We considered three different models that could address existing cosmological tensions. 
While all of these models are consistent with \planck{} data, they exhibit deviations from the \llcdm\ predictions—deviations that are larger than the forecasted \sptnew{} error bars.
This set of alternative models illuminates the potential discovery space of \sptnew{} observations. 
%to test these alternate models, thereby helping us to understand the primordial plasma and early thermal history of the universe, as well as shedding light on the myriad possible forms that dark matter may take on.% \citep[for example,][]{cyrracine21, ferguson22, hughes23}.

These SPT-3G tests of \llcdm\ will begin soon and then improve in precision over the next ~five years. 
%We expect results from the first two years of data from the \mainfield{} to be released later this year. 
%We forecast that this early dataset will offer constraining power on the \llcdm\ model parameters that is comparable to that achieved by the Planck observations with a Figure of Merit (FoM) of 2 relative to \planck{}.
Later this year, we expect to release the first cosmological constraints from the first two years of the \mainfield{} survey observations, and soon afterwards, from the first two years of Ext-4k survey.
We forecast that this early dataset will offer constraining power on the \llcdm\ model parameters that is comparable to that achieved by \planck, improving the \llcdm\ FoM by a factor of 2 (\mainfield) and 13 (Ext-4k) relative to \planck{}. A significant portion of the \extfield{} survey has already been completed and we could expect Ext-10k results as early as sometime in 2025.

\section*{Acknowledgments}
\refresponse{We thank the anonymous referee for their comments which has helped in shaping this manuscript better.}

SR acknowledges support by the Illinois Survey Science Fellowship from the Center for AstroPhysical Surveys at the National Center for Supercomputing Applications.
Work at Argonne National Lab is supported by UChicago Argonne LLC, Operator of Argonne National Laboratory (Argonne). Argonne, a U.S. Department of Energy Office of Science Laboratory, is operated under contract no. DE-AC02-06CH11357. The IAP group acknowledges funding from the European Research Council (ERC) under the European Union’s Horizon 2020 research and innovation programme (grant agreement No 101001897).

The South Pole Telescope program is supported by the National Science Foundation (NSF) through award OPP-1852617. Partial support is also provided by the Kavli Institute of Cosmological Physics at the University of Chicago. 

This work made use of the following computing resources: Illinois Campus Cluster, a computing resource that is operated by the Illinois Campus Cluster Program (ICCP) in conjunction with the National Center for Supercomputing Applications (NCSA) and which is supported by funds from the University of Illinois at Urbana-Champaign; the computational and storage services associated with the Hoffman2 Shared Cluster provided by UCLA Institute for Digital Research and Education's Research Technology Group; the computing resources provided on Crossover, a high-performance computing cluster operated by the Laboratory Computing Resource Center at Argonne National Laboratory; and the National Energy Research Scientific Computing Center, which is supported by the Office of Science of the U.S. Department of Energy under Contract No. DE-AC02-05CH11231.

%--------------------------------------------------------------------------%

\appendix

\section{Derivation of the MUSE covariance matrix}
\label{appendix_muse_cov}
In the context of parameter inferences 
 from CMB data, the observed maps ($x$), do not depend just on cosmological parameters ($\theta$) but also on unobserved latent variables ($z$) such as the unlensed CMB field and the lensing potential field. In such a scenario, the likelihood ($\mathcal{L}(x \mid \theta)$) involves marginalizing the joint likelihood $\mathcal{L}(x, z \mid \theta)$ over the latent space. 
 
\begin{equation}
\mathcal{L}(x \mid \theta)=\int \mathrm{d}^{N}z \mathcal{L}(x, z \mid \theta)=\int \mathrm{d}^{N}z \mathcal{L}(x \mid z, \theta) \mathcal{L}(z \mid \theta),
\end{equation}

where $\mathcal{L}(z \mid \theta)$ represents the likelihood of the unlensed CMB field or the lensing potential $z$, given a set of cosmological parameters $\theta$. For parameter inferences, we are often interested in the gradient of this marginal likelihood, a quantity called the marginal score $s_i$.
 
 \begin{equation}
    s_i(\theta, x) \equiv \frac{d}{d \theta_i} \log \mathcal{L}(x \mid \theta).
\end{equation}

However, performing the integral over the maps of unlensed CMB and lensing potential which contain an order of million pixels is a computationally challenging task and analytic solutions only exist for the simplest cases. MUSE gets around this problem by approximating the marginal score. The key ingredient of the MUSE score is a quantity referred to as $s^{\mathrm{MAP}}$, which is the gradient of the joint likelihood evaluated at the maximum a posteriori (MAP) estimate of the unlensed CMB field and the lensing potential, $\hat{z}_{\theta, x} \equiv \underset{z}{\operatorname{argmax}} \log \mathcal{L}(x, z \mid \theta) $

\begin{equation}
s_i^{\mathrm{MAP}}(\theta, x) \equiv \frac{d}{d \theta_i} \log \mathcal{L}\left(x, \hat{z}_{\theta, x} \mid \theta\right)
\end{equation}

and the MUSE score is defined as  
\begin{equation}
    s_i^{\mathrm{MUSE}}(\theta, x) \equiv s_i^{\mathrm{MAP}}(\theta, x)-\left\langle s_i^{\mathrm{MAP}}\left(\theta, x^{\prime}\right)\right\rangle_{x^{\prime} \sim \mathcal{P}\left(x^{\prime} \mid \theta\right)},
\end{equation}

where the second term is calculated by averaging over simulations $x^{\prime}$ drawn from a probability distribution generated at the specified value of $\theta$. Once we have the MUSE score, we can define an estimator for the bandpowers $\theta$ as the root of the equation $s_i^{\mathrm{MUSE}}(\theta^\mathrm{MUSE}, x) = 0$

and the bandpower covariance can be constructed by two related matrices, J and H defined as:
\begin{equation}
   \begin{aligned}
J_{i j}= & \left\langle s_i^{\mathrm{MAP}}\left(\theta^*, x\right) s_j^{\mathrm{MAP}}\left(\theta^*, x\right)\right\rangle_{x \sim \mathcal{P}\left(x \mid \theta^*\right)} \\
& -\left\langle s_i^{\mathrm{MAP}}\left(\theta^*, x\right)\right\rangle\left\langle s_j^{\mathrm{MAP}}\left(\theta^*, x\right)\right\rangle_{x \sim \mathcal{P}\left(x \mid \theta^*\right)} \\
H_{i j}= & \left.\frac{d}{d \theta_j}\left[\left\langle s_i^{\mathrm{MAP}}\left(\theta^*, x\right)\right\rangle_{x \sim \mathcal{P}(x \mid \theta)}\right]\right|_{\theta=\theta^*}.
\end{aligned}
\end{equation}

$J$ is the covariance of the MAP gradients at the true value of bandpowers $\theta^*$. We use simulations of the data in order to calculate this quantity. $H$ can be thought of as the response matrix of the MAP gradients with respect to small changes in parameters that generate the simulated data. The bandpower covariance is constructed as follows:

\begin{equation}
    \Sigma_{i j}^{\mathrm{MUSE}} \equiv\left\langle\Delta \hat{\theta}_i^{\mathrm{MUSE}} \Delta \hat{\theta}_j^{\mathrm{MUSE}}\right\rangle=\left(H^{-1} J H^{-1\dagger}\right)_{i j}.
    \label{eq:musecov}
\end{equation}

\subsection{Simulations}
\label{appendix_muse_sims}
%We now describe the simulations that are required for calculation of the MUSE covariance matrices. 
We create simulations of the data, which consists of the lensed CMB along with noise as described in Table~\ref{tab_survey_specs}. Utilizing the best-fit \planck{} '15 spectrum, we produce a suite of 1000 flat-sky realizations of unlensed $T, Q,$ and $U$ CMB maps on a 512x512 pixel grid, spanning an area of  $300$ $\sqdeg${}, which are subsequently lensed by the lenseflow algorithm \citep{millea19}. We do not simulate the entire observation patch to limit the computational cost, thereby allowing us to run a large number of simulations. We do not expect this to have a significant impact as lensing is a local operation and the induced deflections are coherent only across a few degrees. Subsequently, we scaled this covariance matrix using an appropriate factor that corresponds to the ratio between our survey area and the simulated patch size. For the purposes of this forecast, we use a simplified version of the data model, which will be used in the actual analysis, whereby we ignore the effect of masking and filtering as is generally done with traditional fisher forecasts. We don't anticipate these to have a significant impact on the parameter constraints. 

The data model used in this work can be written as: 
\[
\begin{aligned}
& d= \mathbb{F} \cdot 
\mathbb{B} \cdot \mathbb{L}(\phi) f+n\\
&f \sim \mathbb{C}_f\left(\boldsymbol{C}_{\ell}^{\mathrm{TT} / \mathrm{TE} / \mathrm{EE}}\right) \\
& \phi \sim \mathbb{C}_\phi\left(\boldsymbol{C}_{\ell}^{\phi \phi}\right) \\
& n \sim \mathbb{C}_n(N_\ell),
\end{aligned}
\]
where, $d$ is the data; $f$ is a Gaussian realization of the unlensed CMB map; $\phi$ is a Gaussian realization of the gravitational lensing potential; $\mathbb{L}(\phi)$ is a linear lensing operator acting on $f$; $\mathbb{B}$ is the instrumental beam function; $\mathbb{F}$ is a mid-pass Fourier filter that cuts off $\ell$s below 300 and above 3500(4000) for $TT, TE(EE)$, and $N_\ell$ is the white+$1/f$ noise power spectrum as defined in Table~\ref{tab_survey_specs}.
\iffalse{
\begin{itemize}
    %\itemsep0em 
    \item $d$ is the data 
    \item $f$ is a Gaussian realization of the unlensed CMB map, 
    \item   $\phi$ is a Gaussian realization of the gravitational lensing potential,
    \item  $L(\phi)$ is a linear lensing operator acting on $f$, 
    \item $\mathbb{B}$ is the instrumental beam function, 
    \item $\mathbb{F}$ is a mid-pass Fourier filter that cuts off $\ell$s below 300 and above 3500(4000) for $TT, TE(EE)$, and
    \item $N_\ell$ is the white+$1/f$ noise power spectrum as defined in Table~\ref{tab_survey_specs}.
\end{itemize}
}\fi

\subsection{Effective lensing reconstruction map noise}
\label{appendix_muse_lensing_noise}

%Before presenting the full forecasting results, we first discuss the effective noise levels of the lensing reconstructions which enter these forecasts. 

The effective lensing reconstruction noise is defined such that for an unbiased estimate of the lensing potential harmonic coefficients, $\hat \phi_{LM}$, the per-mode variance of the residual to the true $\phi$ map is
\begin{equation}
    N_L^{\phi\phi} = \big\langle (\hat \phi_{LM} - \phi_{LM}^{\rm true})^2 \big\rangle.
\end{equation}
For the quadratic estimate, for example, this is given by the sum of the $N^0$, $N^1$, $N^{3/2}$, etc.. noise bias terms \citep{hu02a, madhavacheril20b}. Note that if one has a {\it biased} estimate of the lensing potential, $\bar\phi = A \phi + n$ with some unknown normalization bias, $A$, and effective noise, $n$, the noise power can be written as 
\begin{equation}
    N_L^{\phi\phi} = C_L^{\phi\phi} \left( \frac{1}{\rho_L} - 1 \right),
\end{equation}
where,
\begin{equation}
    \rho_L = \frac{\langle \phi_{LM}^{\rm true} \bar\phi_{LM} \rangle}{\sqrt{\langle \phi_{LM}^{\rm true} \phi_{LM}^{\rm true} \rangle \langle \bar\phi_{LM} \bar\phi_{LM} \rangle}}
    \label{eq:rhoL}
\end{equation}
is the average correlation coefficient between the true and estimated lensing potential. This allows computing the effective noise without ever explicitly debiasing the lensing estimate to remove the bias, $A$. 

The per-mode noise enters the error bars on the estimated lensing power spectrum as, 
\begin{equation}
    \big\langle (\Delta C_L^{\phi\phi})^2 \big\rangle = \frac{2(C_L^{\phi\phi} + N_L^{\phi\phi})^2}{(f_{\rm sky}(2L+1))}.
    \label{eq:fisherCLerrors}
\end{equation} 
 Our MUSE forecasts do not compute an unbiased lensing map estimate, $\hat\phi_{LM}$, nor do they explicitly compute $N_L^{\phi\phi}$ or use Eq.(\ref{eq:fisherCLerrors}). Instead, MUSE directly computes the total posterior bandpower covariance, including signal and noise contributions, using Eq.(\ref{eq:musecov}). It is useful however to explore the effective noise levels which are reached. This can be conveniently performed by computing the averages in $\rho_L$ via Monte Carlo methods, where the biased lensing map estimate, $\bar\phi$, is the MAP estimate of $\phi$, which is a byproduct of the MUSE inference procedure.

\section{Additional details about parameter constraints}
\label{appendix_cosmo_constraints}

In Table~\ref{tab_ext_param_table} we present the constraints on the single-parameter extensions to \llcdm\ from all the \sptnew{} datasets. A comparison of the \allspt{} dataset with \planck{} is presented in Fig.~\ref{fig_single_param_constraints}. In Table~\ref{tab_lcmd_degradation_due_to_ext}, we present the degradation in cosmological parameters when we add one more parameter to the \llcdm\ model. This is the same as the information presented in Fig.~\ref{fig_lcdm_degration_due_to_single_param_ext}. 
\par
\begin{figure*}
    \centering
    \ifdefined\apjformat
        \includegraphics[width=0.9\textwidth, keepaspectratio]{lcdm_constraints_with_and_without_lensing.pdf}
    \else
        \includegraphics[width=0.9\textwidth, keepaspectratio]{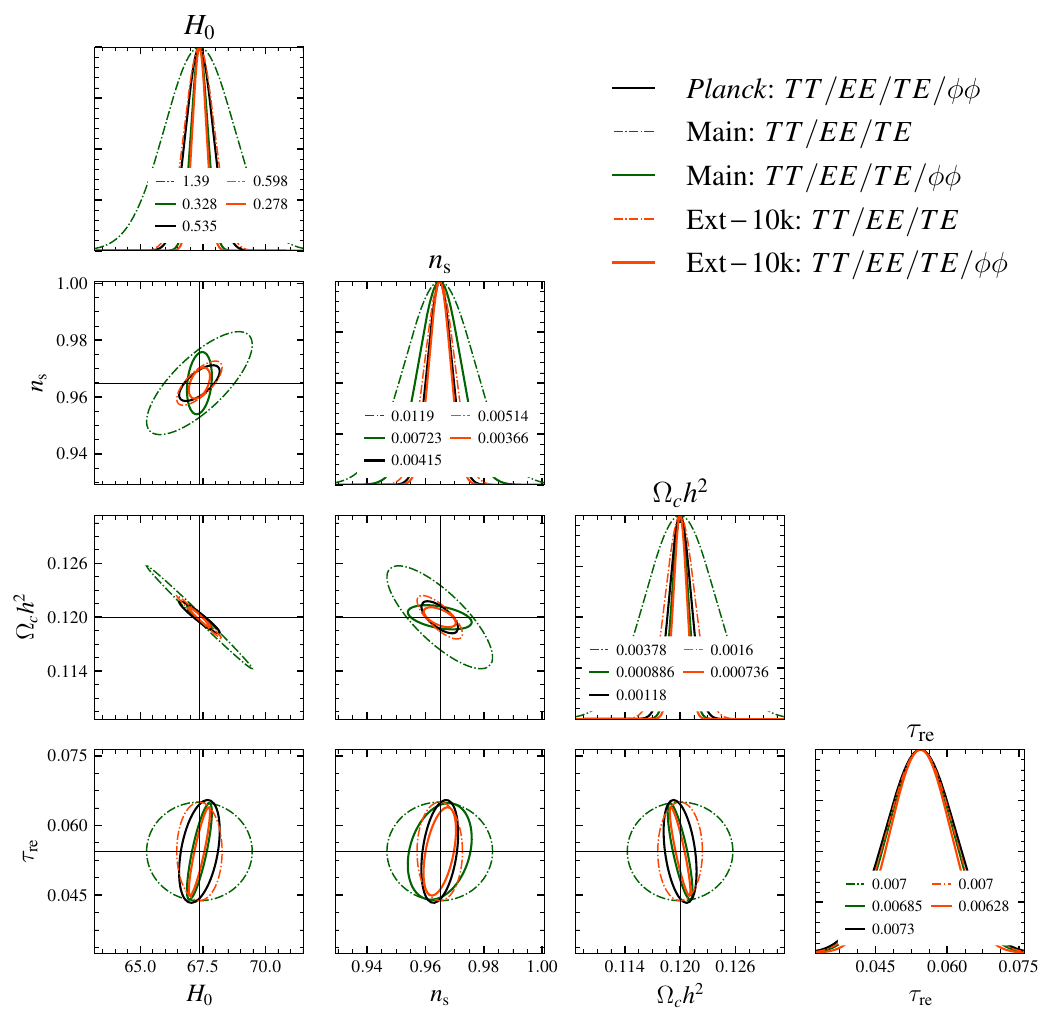}
    \fi
\caption{Marginalized 1-dimensional and 2-dimensional 68\% contours for Hubble constant($H_0$), scalar spectral index ($n_s$), Cold dark matter density ($\omch$), and optical depth ($\taure$), using the MUSE method, for \mainfield{} (dark green), \allspt{} (orange) and \planck{} (black). The dash-dotted curves show the constraints from delensed $TT/TE/EE$ spectra whereas the solid curves also include lensing power spectrum information. The inclusion of the lensing spectrum significantly helps to break the degeneracies especially for the \mainfield{} survey. The $1\sigma$ errors are listed in the legend.}    
\label{fig_lcdm_constraints_different_datasets}
\end{figure*}
\lcdmplussingleparamext{}
\lcdmtabledegradationwithextensions{}

In Fig.~\ref{fig_lcdm_constraints_different_datasets}, we demonstrate the parameter constraints from \planck{} in black, and \mainfield{} in green and \allspt{} in orange using the MUSE method. 
For SPT-3G, we show the constraints that we get from delensed $TT/EE/TE$ CMB-only as the dash-dotted and the ones with the inclusion of lensing $\phi\phi$ as the solid curves. 
It is evident from the figure that the inclusion of lensing, breaks parameter degeneracies, which helps in improving the constraints on many cosmological parameters. 

\bibliography{spt3g_forecasting}{}
\bibliographystyle{aasjournal}

\end{document}